\documentclass[12pt]{article}

\pdfoutput=1

\usepackage{amsmath,amssymb,amsfonts,amscd,mathrsfs}
\usepackage[usenames,dvipsnames]{xcolor}
\usepackage{color}
\definecolor{darkblue}{rgb}{0.1,0.1,.7}
\usepackage[colorlinks, linkcolor=darkblue, citecolor=darkblue, urlcolor=darkblue, linktocpage]{hyperref} 
\usepackage{geometry}
\usepackage{tablefootnote,colortbl}

\usepackage{braket}

\usepackage{lipsum}

\usepackage{empheq}
\newcommand*\widefbox[1]{\fbox{\hspace{.0em}#1\hspace{.0em}}}

\usepackage{cancel}

\usepackage{arydshln}
\usepackage{booktabs,multirow}

\usepackage{tikz-feynman}
 \tikzfeynmanset{compat=1.0.0}
 
 \usepackage[square, comma, compress,numbers]{natbib}

\definecolor{myorange}{RGB}{199,146,32}

\definecolor{Gray1}{gray}{0.97}
\definecolor{Gray2}{gray}{0.9}
\definecolor{LightCyan}{rgb}{0.88,1,1}
\definecolor{blu}{rgb}{0,0,1}

\usepackage{ragged2e}
\usepackage{array}
\newcolumntype{L}[1]{>{\raggedright\let\newline\\\arraybackslash\hspace{0pt}}m{#1}}
\newcolumntype{C}[1]{>{\centering\let\newline\\\arraybackslash\hspace{0pt}}m{#1}}
\newcolumntype{R}[1]{>{\raggedleft\let\newline\\\arraybackslash\hspace{0pt}}m{#1}}

\usepackage{dsfont} 
\usepackage{tcolorbox}
\usepackage{booktabs}

\usepackage{titlesec}
\titleformat*{\section}{\large\bfseries}
\titleformat*{\subsection}{\normalsize\bfseries}
\titleformat*{\subsubsection}{\normalsize\it}
\titleformat*{\paragraph}{\normalsize\bfseries}
\titleformat*{\subparagraph}{\normalsize\bfseries}

\def\bM{{\bf M}}

\newcommand{\reef}[1]{(\ref{#1})}

\def\eps{\epsilon}

\newcommand{\beq}{\begin{equation}} 
\newcommand{\eeq}{\end{equation}}
 
\def\nn{\nonumber}

\def\le{\leqslant}
\def\geq{\geqslant}
\def\leq{\leqslant}
\newcommand{\diffop}[2]{\ifthenelse{\equal{#2}{1}}{\frac{\mrm{d}}{\mrm{d} #1}}{\frac{\mrm{d}^#2}{\mrm{d} #1^#2}}}

\newcommand{\mrm}[1]{{\mathrm #1}}

\usepackage[normalem]{ulem}

\newcommand{\roig}{\color{red}}

\newcommand{\be}{\begin{equation}}
\newcommand{\ee}{\end{equation}}
\def\bea#1\eea{\begin{align}#1\end{align}}

 \def\om{\omega}
  \def\th{\theta}
  
  \def\blu{\color [rgb]{0,.39,.90}}

\newcommand{\dCinc}[4]{
  \begin{minipage}[h]{0.08\linewidth}
    \vspace{.1cm}\begin{tikzpicture}
  [
roundnode/.style={circle, draw=black!60, fill=black!6, very thick, 
  inner sep=1pt,
  text width=3mm},
]
\begin{feynman}[small]
\vertex (x1) at (0,0);
\vertex (x1T) at (-.12,0){\footnotesize$^{{#1}}$};
\vertex (x1aux) at (-.05,-.07);
\vertex (x2) at (.72,0);
\vertex (x2T) at (.7+.14,0){\footnotesize$^{{#3}}$};
\vertex (x2aux) at (.75,-.07);
\vertex (x3) at (0,-.7);
\vertex (x1T) at (-.12,-.8){\footnotesize$^{{#4}}$};
\vertex (x3aux) at (-.05,-.7+.07);
\vertex (x4) at (.72,-.7);
\vertex (x1T) at (.7+.14,-.8){\footnotesize$^{{#2}}$};
\vertex (x4aux) at (.7+.05,-.7+.07);
   \diagram*{
   (x1) -- [very thick, quarter right, looseness=2, Aquamarine] (x2)  ,
   (x3) -- [very thick, quarter left, looseness=2,  Apricot ] (x4)  , 
   };
  \end{feynman}
\end{tikzpicture}
  \end{minipage} 
  }

\newcommand{\dCincDos}[4]{
  \begin{minipage}[h]{0.08\linewidth}
  \vspace{.1cm}\begin{tikzpicture}
  [
roundnode/.style={circle, draw=black!60, fill=black!6, very thick, 
  inner sep=1pt,
  text width=3mm},
]

\begin{feynman}[small]
\vertex (x1) at (0,0);
\vertex (x1T) at (-.12,0){\footnotesize$^{{#1}}$};
\vertex (x1aux) at (-.05,-.07);
\vertex (x2) at (.72,0);
\vertex (x2T) at (.7+.14,0){\footnotesize$^{{#3}}$};
\vertex (x2aux) at (.75,-.07);
\vertex (x3) at (0,-.7);
\vertex (x1T) at (-.12,-.8){\footnotesize$^{{#4}}$};
\vertex (x3aux) at (-.05,-.7+.07);
\vertex (x4) at (.72,-.7);
\vertex (x1T) at (.7+.14,-.8){\footnotesize$^{{#2}}$};
\vertex (x4aux) at (.7+.05,-.7+.07);
\vertex (AUX1) at (0.3,-.4);
\vertex (AUX2) at (.4,-.3);
   \diagram*{
   (x1) --[very thick,  Aquamarine]  (x4)  ,
   (x3) -- [very thick,  Apricot ] (AUX1) ,
   (AUX2) -- [very thick,  Apricot ] (x2)  , 
   };
  \end{feynman}
\end{tikzpicture}
  \end{minipage} 
  }

\newcommand{\dCincTres}[4]{
  \begin{minipage}[h]{0.09\linewidth}
    \vspace{.1cm}\begin{tikzpicture}
  [
roundnode/.style={circle, draw=black!60, fill=black!6, very thick, 
  inner sep=1pt,
  text width=3mm},
]
\begin{feynman}[small]
\vertex (x1) at (0,0);
\vertex (x1T) at (-.12,0){\footnotesize$^{{#1}}$};
\vertex (x1aux) at (-.05,-.07);
\vertex (x2) at (.72,0);
\vertex (x2T) at (.7+.14,0){\footnotesize$^{{#3}}$};
\vertex (x2aux) at (.75,-.07);
\vertex (x3) at (0,-.7);
\vertex (x1T) at (-.12,-.8){\footnotesize$^{{#4}}$};
\vertex (x3aux) at (-.05,-.7+.07);
\vertex (x4) at (.72,-.7);
\vertex (x1T) at (.7+.14,-.8){\footnotesize$^{{#2}}$};
\vertex (x4aux) at (.7+.05,-.7+.07);
   \diagram*{
   (x1) -- [very thick, quarter left, looseness=2, Apricot ] (x3)  ,
   (x2) -- [very thick, quarter right, looseness=2, Aquamarine] (x4)  , 
   };
  \end{feynman}
\end{tikzpicture}
  \end{minipage} 
  }

\usepackage{accents}
\newlength{\dhatheight}

\newcommand{\im}{\text{Im}}
\newcommand{\re}{\text{Re}}

\numberwithin{equation}{section}
\usepackage{tocloft}
\begin{document}

\vspace*{-.6in} \thispagestyle{empty}
\begin{flushright}
\end{flushright}
\vspace{1cm} {\Large
\begin{center}
\textbf{
Dual EFT Bootstrap:    QCD flux tubes
 }
\end{center}}
\vspace{1cm}
\begin{center}

{\bf  Joan Elias Mir\'o$^{a}$, Andrea Guerrieri$^{b}$} \\[1cm]  {$^a$ The Abdus Salam ICTP,    Strada Costiera 11, 34135, Trieste, Italy  \\ $^b$    School of Physics and Astronomy, Tel Aviv University, Ramat Aviv 69978, Israel  \\ }
\vspace{1cm}

\abstract{

We develop a bootstrap approach to Effective Field Theories (EFTs) based on the concept of duality in optimisation theory. 
As a first application, we consider the fascinating set of EFTs for confining flux tubes. 
The outcome of our analysis are optimal bounds on the  
scattering amplitude of Goldstone excitations of the flux tube, which in turn translate into bounds on 
the Wilson coefficients of the EFT action. 
Finally, we comment on how our approach compares to EFT positivity bounds. 
}

\vspace{3cm}
\end{center}

 \vfill
 {
  \flushright
 \today 
}

\newpage 

\setcounter{tocdepth}{1}

{
\tableofcontents
}

\section{Introduction and motivation}

It is widely appreciated that the paradigm of Effective Field Theory (EFT) is very much universal.
However, despite the wide range of  application and flexibility of EFTs, the principles of unitary evolution and causality imply very interesting bounds on the space of feasible EFTs, i.e. EFTs with a putative UV completion. 
A classic example is provided by the positivity bounds: while a priori Wilson coefficients can take any real value,   positivity  of the two-to-two forward scattering amplitude $\im M>0$  implies that various Wilson coefficients  are positive~\cite{Adams:2006sv}.
Many  works have exploited positivity, including: the  original studies in the context of the chiral Lagrangian \cite{Pham:1985cr,Pennington:1994kc,Ananthanarayan:1994hf}, many interesting  applications on RG-flows and the phenomenology of EFT interactions, see e.g.  \cite{Komargodski:2011vj,Bellazzini:2016xrt,Cheung:2016yqr,Luty:2012ww,Distler:2006if,Englert:2019zmt,Bellazzini:2017fep,Alberte:2020bdz,Bellazzini:2019bzh,Gu:2020ldn,deRham:2018qqo}, as well as  new developments \cite{Arkani-Hamed:2020blm,Green:2019tpt,Bellazzini:2020cot,Tolley:2020gtv,Caron-Huot:2020cmc,Caron-Huot:2021rmr,Bern:2021ppb,Chiang:2021ziz}.

Recent progress on the S-matrix bootstrap programme \cite{Paulos:2016fap,Paulos:2016but,Paulos:2017fhb,Guerrieri:2018uew,Homrich:2019cbt,Karateev:2019ymz,Hebbar:2020ukp,Correia:2020xtr}
has triggered a revision of the space of feasible  EFTs,   with  applications to the  EFT of: the QCD string~\cite{EliasMiro:2019kyf}, pions~\cite{Guerrieri:2020bto,Bose:2020shm,Bose:2020cod} and supergravity \cite{Guerrieri:2021ivu}.
At this point a small digression is in order. Say -- we are interested in the problem of  finding the minimal value of a particular Wilson coefficient in an EFT action\footnote{or  the minimal value of the closely related Low Energy Constant   in the scattering S-matrix.}. 
We can view this task as an optimisation problem
subject to the constraints dictated by unitarity and causality.  
There are two possible logical routes to approach the problem: {\bf a)} search in the space of all physical  theories, and pick the one which achieves the smallest Wilson coefficient (Primal S-matrix bootstrap); or, {\bf b)} exclude all  the values of the Wilson coefficient that are incompatible with either unitarity or causality, and claim a bound on the minimal Wilson coefficient (Dual S-matrix bootstrap).
\be
\includegraphics[width=0.49\textwidth]{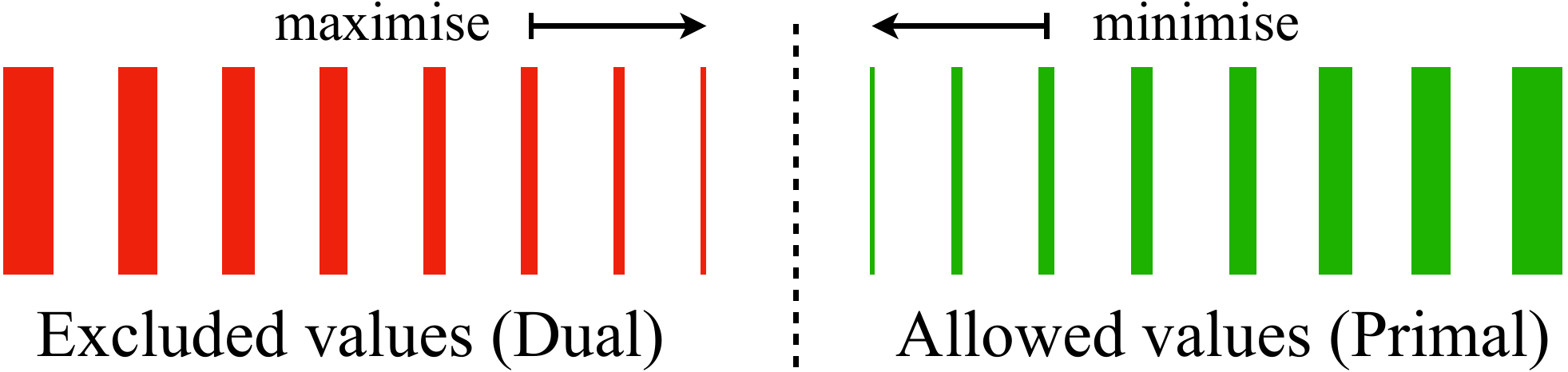} \nonumber
\ee
When the minimal value found from the Primal approach and the maximal of  Dual approach touch each other, indicated with a dashed line above, the duality gap is closed. 
The concept of duality in optimisation theory has been successfully applied to bound the space of $O(N)$  models \cite{Cordova:2019lot} and  the couplings of bound states \cite{Guerrieri:2020kcs} in two spacetime dimensions,   and  quartic couplings in four spacetime dimensions~\cite{He:2021eqn,GS2021}.~\footnote{The primal bootstrap approach to these problems was studied in \cite{He:2018uxa,Cordova:2018uop,Paulos:2018fym,Kruczenski:2020ujw}, as well as \cite{Paulos:2016fap,Paulos:2016but,Paulos:2017fhb,Homrich:2019cbt}.}
The logic of the dual S-Matrix bootstrap approach  resembles that of the CFT bootstrap~\cite{Rattazzi:2008pe}, 
were kinks and island are found \cite{ElShowk:2012ht,El-Showk:2014dwa,Kos:2016ysd} after excluding allowed values of the operator's scaling dimensions.

In this work we  will show how to optimally bound, using a dual formulation, the allowed values of  Wilson coefficients or Low Energy Constants (LECs).
In order to do so we will focus our attention on   the EFT of confining flux tubes~\cite{Luscher:1980ac,Luscher:1980fr},  see also   \cite{Dubovsky:2012sh,Aharony:2013ipa,Caselle:2021eir} and references there in. 
This system is very fascinating per se, describing the long strings of confining  three and four-dimensional theories~\cite{Luscher:1980ac,Luscher:2004ib}, 
and features an interesting phenomenology~\cite{Athenodorou:2010cs,Dubovsky:2013gi}.
It also provides a simplified setting to test our ideas for bounding the space of EFTs. 
At low energies, the flux tube can be described by a two-dimensional action
 given by
\be
A=\int d^2\sigma \sqrt{-h}\left[\ell_s^{-2}+R(h)+K^2+\ell_s^2 g_1(K^\mu_{\alpha\beta} K_\mu^{ \alpha\beta})^2+\ell_s^2 g_2 \, K_\nu^{\alpha\beta}K^\mu_{\alpha\beta} K_\mu^{ \sigma\rho}K^\nu_{ \sigma\rho} + O(\partial^8)\right] \label{act1} \, . 
\ee
The action is build out of the fields $X^\mu(\sigma)$, describing the embedding coordinates of the world-sheet in spacetime. 
In the rest of the paper we will work in units set by the string length $\ell_s=1$, and in the static gauge $X^\mu(\sigma)=(\sigma^\alpha, X^a)$, where $a=1,\dots, D-2$. The action is invariant under the $SO(D-2)$ transverse rotations, such that $X^a$ carries a vector (or flavour) index, and the Poincar\'e sub-group on the world-sheet $ISO(1,1)$.
The goldstone particles created by the fields $X^a$ are called branons. 

At low energy, the leading piece in the  action is  the Nambu-Goto (NG) interaction $\sqrt{-h}=\sqrt{-\text{det}\partial_\alpha X^\mu \partial_\beta X^\nu\eta_{\mu\nu}}$. 
On top of the NG interaction, and following the usual EFT logic, we include in the action any RG-irrelevant interactions that are allowed by the symmetries. Thus we include invariants build out of the intrinsic metric  $h_{\alpha\beta}=\partial_\alpha X^\mu \partial_\beta X^\nu\eta_{\mu\nu}$ (like for instance the Ricci curvature scalar $R(h)$) and the extrinsic curvature $K_{\alpha\beta}^\mu= \nabla_\alpha\partial_\beta X^\mu$.
It turns out  however that $R(h)=0$ in two spacetime dimensions and that $K^2$ vanishes  being  proportional to the equations of motion. 
 This is known as low energy universality~\cite{Luscher:2004ib,Aharony:2009gg,Aharony:2010cx,Aharony:2011gb,Dubovsky:2012sh,Aharony:2013ipa}.

The leading deviation from the universal NG interaction, which is sensitive to the underlying confining dynamics, arrises at order $O(K^4)=O(\partial^6)$,
parametrised by $g_1$ and $g_2$ in the action \reef{act1}.
 In this work we will bound the values of these non-universal interactions. 
In order to do so, we will use the world-sheet S-matrix, describing the scattering  of the branons $X^a$. 
In particular we will need the two-to-two S-matrix, which is given by~\cite{Conkey:2016qju,EliasMiro:2019kyf}
\bea
	2\delta_{sym} &=  \frac{s}{4}   + \alpha_2  s^2 + 
	\alpha_3 s^3+O(s^4) \, ,  \nonumber \\
	2\delta_{anti}  &=  \frac{s}{4}- \alpha_2 s^2 + (\alpha_3 {+}2\beta_3) s^3 +O(s^4)  \, ,  \nonumber \\
	2\delta_{sing} &= \frac{s}{4}- (D{-}3)\alpha_2 s^2 +(\alpha_3 {-}(D{-}2)\beta_3) s^3 +O(s^4)  \, ,  \label{d4lecs}
\eea
where $\alpha_2=\frac{D-26}{384\pi}$ is a universal one-loop contribution~\cite{Polchinski:1991ax,Dubovsky:2012sh}, and  we are using the conventional definition for the S-matrix $S_I(s)=\exp 2 \delta_I(s)$ were $I=sym ,anti,sing$.
While further details are given in sec.~\ref{D4dual}, note that 
thanks  to the $SO(D-2)$ symmetry $X^a\rightarrow R_b^a X_b$,  the two-to-two scattering can proceed in three channels (symmetric, antisymmetric and singlet), corresponding to the three irreducible representations of the incoming $SO(D-2)$ vectors $X^a+X^b\rightarrow X^c +X^d$.~\footnote{
Also recall that, after factoring out the usual delta function of total two-momenta conservation, the  $S_I$'s depends only on the   Mandelstam variable $s=(p_1^\mu+p_2^\mu)^2$ because  in two spacetime dimensions there is no scattering angle (i.e. $t=0$) and because of the Mandelstam relation $s+t+u=0$.  }

The non-universal interactions in \reef{act1} are parametrised in \reef{d4lecs}  through $\{\alpha_3,\beta_3\}$.~\footnote{In particular $\{\alpha_3,\beta_3\}=\{2g_1+3g_2,-2g_1-g_2\}/8$, although the precise matching is not important for our current purposes.}
Our bounds on  the S-matrix parameters  translate into bounds on the energy levels computed in  \cite{EliasMiro:2019kyf}, which in turn can be compared against lattice Monte Carlo (MC) simulations of four-dimensional Yang-Mills. 
The worldsheet S-matrix approach to the QCD flux tube and its interplay with lattice  MC data was pioneered in \cite{Dubovsky:2013gi,Dubovsky:2014fma}; 
see also \cite{Teper:2009uf} for a nice review of flux tubes from a lattice MC viewpoint. 

In section~\ref{D3dual} we introduce the formalism of dual EFT bootstrap. 
In order to do so we start discussing the  flux tube in $D=3$ bulk spacetime dimensions, which has an additional pedagogic value because it is a simpler problem. 
In section~\ref{D4dual}  we generalize the discussion to flux tubes in general $D>3$ target spacetime dimensions and present the bounds on $\{\alpha_3,\beta_3\}$.
See table~\ref{tabbb} for a summary of what we know on the bootstrap approach to the EFT of flux tubes. 
A nice feature of the bootstrap approach is that it delivers the S-matrix saturating the bounds. 
In section~\ref{cps} we discuss the phenomenology of these dual S-matrices. 
In \ref{cok} we conclude and  discuss the interplay  of positivity v.s. bootstrap. 
Finally, appendices \ref{ndp}, \ref{ggg} and  \ref{bonus} are dedicated to give further details on the numerics, on the generalisation of $D=3$ and $D\geq4$ analysis, respectively.

{ \renewcommand{\arraystretch}{1.3} \renewcommand\tabcolsep{6.4pt}
\begin{table}[t]
\begin{center}
\begin{tabular}{L{4.5cm} L{2.9cm}  L{2.9cm}   }  
\toprule  
 & $\bf{D=3 }$ & $\bf{D\geq 4}$    \\ \rowcolor{Gray1}
\midrule
      \bf{Primal formulation}    &   \checkmark~\cite{EliasMiro:2019kyf}   &      \checkmark~\cite{EliasMiro:2019kyf}    \\  
 \bf{Dual formulation}              &    \checkmark~\S~\ref{D3dual}   &     \checkmark~\S~\ref{D4dual}   \\   \rowcolor{Gray1}
 
      \bf{Analytical solution}    &   \checkmark~\S~\ref{analit}~and~\cite{EliasMiro:2019kyf}    &      unknown to us    \\  
   \bottomrule
\end{tabular}
\caption[Works]{ 
Optimization of low energy constants (LECs) of the flux tube EFTs.  
     \label{tabbb} }
\end{center} \vspace{-.5cm}
\end{table}
}

\section{Dual optimisation of Wilson coefficients}
\label{D3dual}

In order to develop the theory of dual optimisation of Wilson coefficients, we start by analysing the scattering 
of a single-flavour gapless branon, a.k.a. $D=3$ flux tubes.
The three processes in \reef{d4lecs} reduce to a single channel $S(s)=e^{2i \delta(s)}$, with $\delta=\delta_\text{sing}$, and a single non-universal parameter is needed at $O(s^3)$, $\alpha_3-\beta_3\equiv\gamma_3$. 

The S-matrix is the boundary value of the function $S(s)$  which is analytic in the upper half plane (UHP) of the complexified Mandelstam variable $s=(p_1+p_2)^2$. The value of the function at  specular points with respect to  the imaginary axis are related by complex conjugation  \be
S(-z^*)=S^*(z) \, , \label{xsing1}
\ee 
as a consequence of   crossing-symmetry $S(-z)=S(z)$ and real-analyticity  $S(z^*)=S^*(z)$.
A nice discussion of the properties of the  scattering S-matrix of massless particles in two spacetime dimensions can be  found in \cite{Zamolodchikov:1991vx}.
Since $S(z)$ is the expectation value of a unitary operator it satisfies
\be
|S(s)|\leq 1 \quad \text{for}\quad s\in(0,\infty) \, , \label{unit}
\ee 
i.e. for physical values of the Mandelstam variable $s$.

The spontaneously broken Poincar\'e invariance strongly constrains the low energy behaviour of the two-to-two phase shift~\cite{Chen:2018keo,EliasMiro:2019kyf}~\footnote{The phase shift is real up to $O(s^8)$ when $2\to 4$ particle production processes kick in.}
\be
2\delta(s)  = \frac{s}{4}+ \gamma_3 s^3 + \gamma_5 s^5+ \gamma_7 s^7+ O(s^8) \, . \label{lecs1}
\ee
The coefficients $\gamma_i$  are tuneable real parameters of the low energy EFT,  that should be fitted to  low energy experimental  data (or to MC lattice simulations data \cite{Athenodorou:2016kpd}), and whose precise values depend on the details of a putative UV completion. 
However, the $\gamma_i$'s  do not take  arbitrary real values but instead satisfy  sharp bounds that follow as a consequence of  unitary \reef{unit}, crossing and real-analyticity \reef{xsing1}.

\subsection{Primal optimisation problem}

To be concrete and explain in detail the general strategy of dual optimization for Wilson coefficients, in the rest of the  section we will address the specific problem of finding the \emph{minimal value of $\gamma_3$}.

The first simple strategy to approach this problem is based on the direct numerical optimisation. 
In a nutshell,  one introduces an ansatz for the S-matrix which encodes automatically the analytical and crossing properties \reef{xsing1}, and the low energy expansion \reef{lecs1}. 
This is for instance achieved by
 \bea
 S_\text{ansatz}(\chi)=\sum_{n=0}^{n_\text{max}}  \alpha_n \, (\chi - 1)^n \quad \text{with} \quad \chi(s) = \frac{s-i}{s+i} \, ,
 \label{ans1}
 \eea
with the parameters  $\{  \alpha_0,   \alpha_1,   \alpha_2,   \alpha_3\}$ fixed to match the low energy expansion $S_\text{ansatz}(\chi(s))=\exp[i2 \delta(s)]+O(s^4)$ \reef{lecs1}. 
Next, we minimize $\gamma_3$ varying over the remaining $\alpha_{n\geq 4}$ subject to the unitary constraint~\reef{unit}.
This basic logic can be generalised to higher dimensions 
and has been successfully used to explore the extremal values of  the LECs of pion physics \cite{Guerrieri:2020bto} and supergravity \cite{Guerrieri:2021ivu}.

In the case at hand however,  
an analytical  solution  was found in \cite{EliasMiro:2019kyf}
\be
\gamma_3 \geq -\frac{1}{768}.
\label{b1}
\ee
The proof presented there is based on the Schwarz-Pick inequality.~\footnote{This analytic result fits in the general \emph{geometric function theory} recently reviewed in \cite{Haldar:2021rri} and generalised to other interesting physical examples.}
Consider the  following function of $z$ constructed out of a physical $S$-matrix $S(z)$
\beq
S^{(1)}(z|w) \equiv \frac{S(z)- S(w) }{1- S(z) \overline{S(w)}} \Big/\frac{z-w}{z - \overline w} \, ,    \label{SP1}
\eeq
where $w$ is an arbitrary point in the upper half plane.
Next, note that   (as a holomorphic function of $z$) this function  has no singularities in the upper half plane and by unitarity is  bounded by 1 for $z$ on the real line, $|S^{(1)}(s|w)| \le 1$ for $s\in \mathbb{R}$. Then, by the maximum modulus principle, $S^{(1)}(z|w)$  is bounded everywhere on the upper half plane
\be
|S^{(1)}(z|w)|  \le 1 \quad \text{for}\quad z \in\text{UHP}  \, .
\ee
The last equation is the content of the Schwarz-Pick theorem.
Finally, inserting the low energy expansion \reef{lecs1}   in the Schwarz-Pick function \eqref{SP1}
and expanding  for small and imaginary $z$ and $w$,  
\be
S^{(1)}(i x |i y)=-1+\left(\frac{1}{96}+8 \gamma_3\right) x\,y+\dots \geq -1 \, ,  \label{limit1}
\ee  leads to \reef{b1}. 
The logic flow just presented can be recursed over, i.e. one can build a $S^{(2)}$ function out of $S^{(1)}$ to bound $\gamma_5$, and so on.~\footnote{While further details are provided in \cite{EliasMiro:2019kyf}, we recall that the Schwarz-Pick bounds are saturated by products of Castillejo-Dalitz-Dyson (CDD) factors (known as Blaschke products in complex analysis   literature). Indeed, it is straightforward to check that the first Schwarz-Pick bound  \reef{b1}  is saturated by  
 $
S_\text{opt}(z)= \frac{i 8-z}{i8+z} 
 $.
The later function is associated  (i.e. equal modulo a sign) to the goldstino S-matrix that describes the flow  from the  Tricritical  to the Critical Ising fixed points~\cite{Zamolodchikov:1991vx}.}

In the next section we will derive an alternative proof of this bound based on duality in optimization theory.~\footnote{A nice textbook is for instance \cite{cvx}.}
We will work out in detail the dual formulation of the primal problem we just solved  generalizing the procedure introduced in \cite{Guerrieri:2020kcs} for gapped theories, and highlight the various novel aspects related to gapless systems. This will clear the way for section \ref{D4dual}  where we will be able to use the dual formulation to bootstrap max/min values of the Wilson coefficients in situations  where no analytical solution is known.

\subsection{Dual optimisation problem}
\label{analit}

To derive the dual problem it is convenient to  formulate  the primal approach   in terms of the two to two scattering amplitudes and the associated dispersion relations. The parameter $\gamma_3$ appears in the low energy expansion of the flux tube amplitude through \reef{lecs1}, i.e.
\be
M^\text{FT}(s)=\frac{s^2}{2}+\frac{i s^3}{16}-   \vphantom{\int_0^\infty } \left(\frac{1}{192}-2  \gamma_3\right) s^4+O\left(s^5\right) \, .  \label{ampft}
\ee
The amplitude   $M^\text{FT}(s)$ is subject to  unitary \reef{unit}, and real-analyticity and crossing  \reef{xsing1}.~\footnote{Recall that 
 $ i M(s)\equiv  2s (S(s)-1)$, where the factor $s$ arises as a Jacobian in the relation of the identity operator of the S-matrix  $\hat S=x  \mathds{1} S(s)$, where $\mathds{1} = (2\pi)^2 s (\delta(p_1-p_3)\delta(p_1-p_4)+(3\leftrightarrow 4))$, and the two-momentum conservation delta in the  interacting scattering  amplitude  $\hat  M= (2\pi)^2\delta^{(2)}(k_1^\mu+k_2^\mu-k_3^\mu-k_4^\mu)M(s)$, with $k_i^\mu=(|p_i|,p_i)$.}
We write the upper index in $M^\text{FT}$ to distinguish an arbitrary amplitude from  the actual flux-tube amplitude. 
We formulate the primal optimization problem writing all the constraints explicitly: 
\begin{subequations}
\begin{empheq}[box=\widefbox]{align}
\nonumber\\[-.1cm]
&  \text{\hypertarget{pp1}{Primal Problem I}:}  \nonumber\\[.1cm]
& \text{Minimise }  \gamma_3 \text{ varying } M(s)\text{  constrained by}   \nonumber  \\
 &  \circ \,    U(s)   \equiv     2\, \text{Im} M(s) - \frac{1}{2s} |M(s)|^2  \geq 0 \  \text{ for } s>0 \,  ,  \label{unit2}  \\
  &   \circ \,  
 \text{Disp}(s)  \equiv  \frac{1}{2}\text{Re}M(s)  - \frac{1}{2\pi  }  \int_0^\infty  \frac{s^2}{z^2} \text{Im}M(z)\left(    p.v.\frac{1}{z-s} +\frac{1}{s+z}\right) dz =0 \ \text{ for } s>0 \, ,   \label{anal2} \\ 
&  \circ \,   a_2(0)\equiv  \frac{2}{\pi} \int_0^\infty \frac{ \text{Im}M(z)}{z^3} dz = c_2  \ \text{ with } \ c_2=\frac{1}{2} \, ,  \label{lec1} \\
&  \circ \,     a_3(0)\equiv - \frac{2}{\pi} \int_0^\infty \frac{ \text{Re} M(z)-c_2 z^2}{z^4} dz = c_3  \ \text{ with } \  c_3=\frac{1}{16}  \, , \label{lec2} \\
&  \circ \,     a_4(0)\equiv - \frac{2}{\pi} \int_0^\infty \frac{ \text{Im} M(z)-c_3 z^3}{z^5} dz = c_4  \ \text{ with } \ c_4=\frac{1}{192}-2  \gamma_3  \, . \label{lec3} 
\\[-.2cm] \nonumber
\end{empheq}
\end{subequations}

Note that  the constraint \reef{anal2} is satisfied if and only if $M(s)$ 
 is an analytic function in the UHP, which satisfies  $M(-s^*)=M^*(s)$ and unitarity \reef{unit2}.
 To prove the last statement we start with  the following contour integral
 \be
 M(s) =\frac{1}{2\pi i }\oint_{C(s)} \frac{s^2}{z^2}\frac{M(z)}{z-s} dz
 \ee
 that encircles counter-clockwise an arbitrary point $s\in\text{UHP}$.
We introduced a double subtraction to take into account the most general behaviour at infinity compatible with unitarity \reef{unit2}.
 Next we blow up the contour,  use $M^*(z)=M(-z^{*})$ and take $s$ real:
 \be
 M(s)=    \frac{1}{2\pi i} \int_0^\infty \frac{s^2}{z^2} \left( \frac{M(z)}{z-s-i0}-\frac{M^*(z)}{z+s}  \right)dz \, ,  \label{blow}
 \ee
 where we kept a small positive imaginary part in $s+i0$ when needed. 
The double pole at $z=0$ does not pick any residue in virtue of the soft low energy behaviour of the branon amplitude \eqref{ampft}.
Taking the real part of the last equation, and using the Cauchy principal value ($p.v.$), we get   \reef{anal2}.
 
\begin{figure}[t]
\centering
\includegraphics[width=0.4\textwidth]{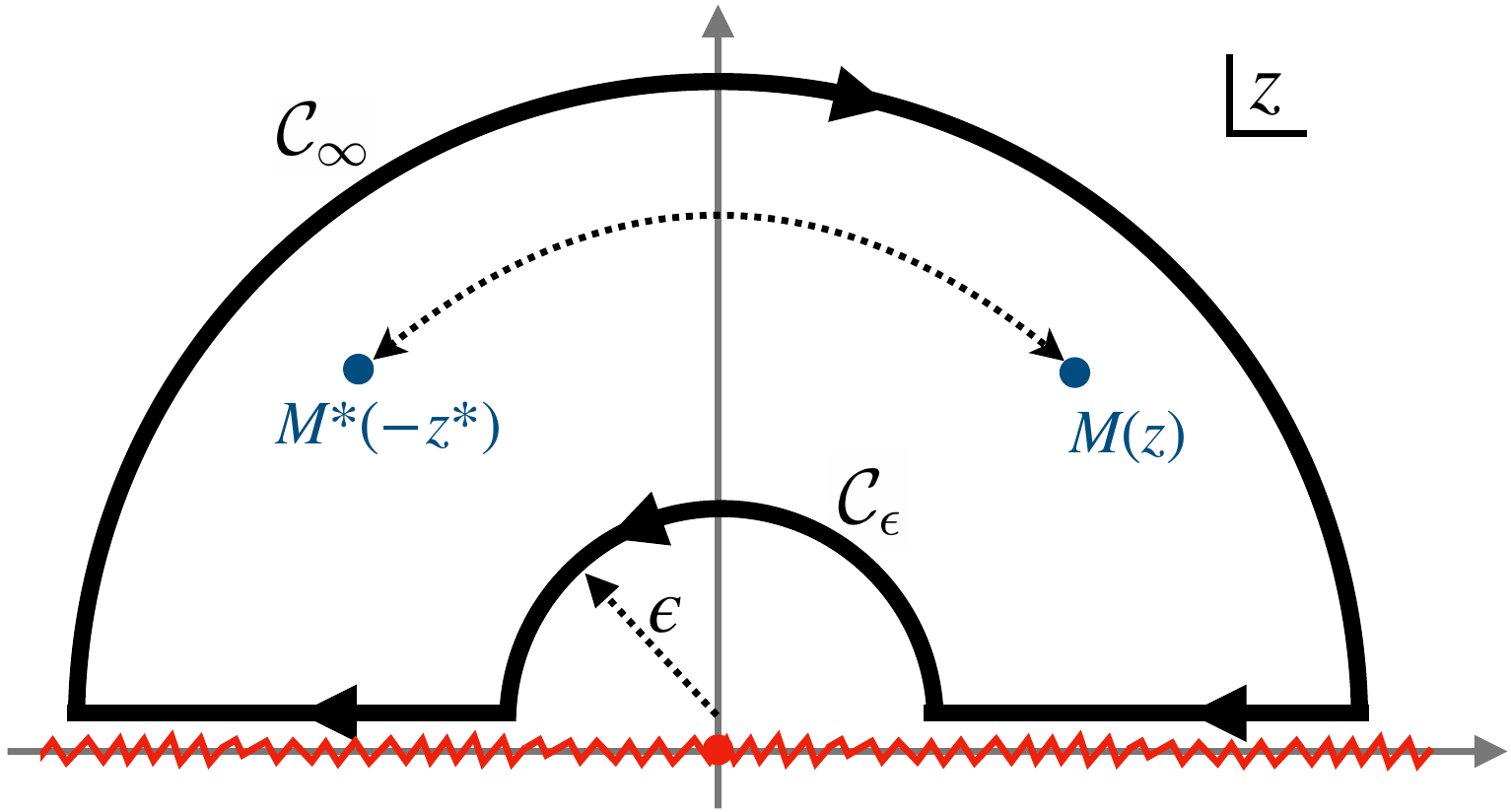}
\caption{Contour of integration used to relate \reef{arcs} with \reef{lec1}-\reef{lec2}. }
\label{contour}
\end{figure}

 Regarding the low energy constraints (\ref{lec1}-\ref{lec2}),  when analyticity and crossing \reef{anal2} are satisfied, we can deform the integration contours in (\ref{lec1}-\ref{lec2}) and write
\be
a_n(\eps ) = (- 1)^{n} \int_{{\cal C}_\eps} \frac{  M(z)- \sum_{m=2}^{n-1}(-1)^{m+1} (-i z)^m c_m}{\pi (-i z)^{n+1}} dz  \,  , \label{arcs} 
\ee
where ${\cal C}_\eps$ is a counter-clockwise semicircle contour in the UHP and centred around $z=0$, see fig.~\ref{contour}. For $\eps\ll 1$, the integral  in \reef{arcs} can be evaluated using the low energy expansion in  \reef{ampft} 
   \be
a_n(\eps)=c_n+ O(\eps) \, ,  \label{tt1}
\ee
with  the $c_n$'s fixed by matching the function $M^\text{FT}(s)=\sum_{m=2}^9 (-1)^{m+1} (-i s)^mc_m+O(s^5)$ with  low energy expansion \reef{ampft}.
In particular, we have 
 $
a_4(\eps)= 1/192 - 2\gamma_3+O(\eps)$ when evaluating \reef{arcs} with $M^\text{FT}$.

Similar variables to $a_n(\eps)$ where recently used in \cite{Bellazzini:2020cot}, there named arcs,  to study the  positivity constraints of operator's Wilson coefficients along the Rernormalization Group flow. In this work, thanks to our knowledge of the low energy expansion \reef{lecs1}, we have introduced subtractions in the definition of $a_n(\eps)$ such that we get \reef{tt1}.

The   formulation of  \hyperlink{pp1}{Primal Problem I} in terms of dispersion relations pays off now because we can encode all the constraints 
in the following quadratic Lagrangian functional
\be
L({\bf M};  \Lambda )  =
2\gamma_3  +
 \underbrace{ \vphantom{\int_0^\infty }  \lambda_2 (a_2(0)- c_2) + \lambda_3 (a_3(0)-c_3)+ \lambda_4 (a_4(0)-c_4)}_\text{low energy constants constraints} +
 \underbrace{ \int_0^\infty \om(z) \text{Disp}(z)  -  \mu(z) U(z)}_\text{analyticity and unitarity constraints} \, , \label{lag1}
\ee
where $\gamma_3$ is our optimisation goal, and we have introduced a dual variable for each constraint in (\ref{unit2}-\ref{lec3}).
${\bf M}$ and $\Lambda$ collectively denotes all the primal and dual variables respectively  
\be
\bM=\{\text{Re}M(z),\text{Im}M(z), \gamma_3 \} \ , \quad \Lambda= \{\lambda_2,\, \lambda_3,\, \lambda_4,\, \om(z),\, \mu(z)\} \, . 
\ee We stress that $\lambda_2,\lambda_3,\lambda_4\in\mathds{R}$, and  $\om(z)$ and $\mu(z)$ are real functions defined for $z>0$.
Is is useful to think of $M(z)$, $\om(z)$ and $\mu(z)$ as local fields of a field-theory.
While the $M(z)$ variables -- one for each point in the real positive line $z\in\mathds{R}^+$ -- are a priori arbitrary, it  turns out that for $M(z)=M^\text{FT}(z)$ in \reef{ampft} the  \emph{low energy constants  constraints} in \reef{lag1} are finite.

At this point we are ready to introduce the dual functional
\be
d(\Lambda)   \equiv \underset{{\bM}}{\phantom{p}\text{inf}\phantom{p}}    L({\bf M}; \Lambda)  \, ,
\ee
obtained by minimising  the Lagrangian w.r.t. varying $\bf M$.
It turns out that $d(\Lambda)$ satisfies the following inequalities
\be
d (\Lambda) 
\, \leq  \,  \underset{ \Lambda }{\text{sup}} \,  d(\Lambda) 
\, =  \,  \underset{ \Lambda }{\text{sup}}        \underset{\bM}{\phantom{p}\text{inf}\phantom{p}}    L({\bM}; \Lambda) 
 \, \leq   \,  \underset{\bM}{\phantom{p}\text{inf}\phantom{p}}    \underset{\Lambda }{\text{sup}}  \,   L(\bM; \Lambda)  
\,  = \,   2\gamma_3^* \label{duality1}
\ee
where  $\gamma_3^*$ is the solution to   \hyperlink{pp1}{Primal Problem I}.
Indeed, the second inequality follows from the Min-Max theorem,  switching the order of the action of sup(remum) and inf(imum). 
The last equality holds because  $\text{sup}_{ \Lambda}  \,   L(\bM ; \,   \Lambda)=+\infty$ if any of the constraints is not satisfied, while 
$\text{sup}_{  \Lambda}   \,   L (\bM ; \, \Lambda )=2 \gamma_3$ if $M$ is feasible, i.e. if all the constraints are satisfied. 
Eq.~\reef{duality1} provides the basis for formulating 
\begin{subequations}
\begin{empheq}[box=\widefbox]{align}
\nonumber\\[-.1cm]
&  \text{\hypertarget{pp2}{Dual Problem I}:}  \nonumber\\[.1cm]
& \text{Maximize} \  d(\Lambda)\text{ varying }  \Lambda=\{\om(z),\,  \mu(z), \, \lambda_2, \, \lambda_3,\, \lambda_4 \}  ,\,    \text{constrained by } 
\mu(z)  > 0  \, . 
\end{empheq}
\end{subequations}

The general logic to get  to formulate \hyperlink{pp2}{Dual Problem I} parallels that of \cite{Guerrieri:2020kcs}.  
Next we will  solve \hyperlink{pp2}{Dual Problem I} and find novel aspects particular to bootstrapping EFTs. 
In doing so we will   show that indeed the solution of \hyperlink{pp2}{Dual Problem I} and  \hyperlink{pp1}{Primal Problem I} coincide. 

In order to find $d(\Lambda)$ we will use the Euler-Lagrange equations of motion (e.o.m.) applied to  \reef{lag1}.
Before doing that, note that the Lagrangian \reef{lag1} is non-local in  $M(z)$ because  it appears integrated over the real line in $\text{Disp}(z)$, defined in \reef{anal2}. It is useful to introduce   the function
\be
W(z^\prime) = \frac{1}{2\pi} \int_0^\infty  \om(z)   \, \frac{z^2}{z^{\prime 2}} \left(  \frac{1}{z-z^\prime -i 0}- \frac{1}{z+z^\prime-i 0}  \right) dz \, . \label{w1}
\ee
because in terms of $W$  the Lagrangian is an integral of a local density. Indeed, using  
 $
\int_0^\infty \om(z) \text{Disp}(z)$ $=\int_0^\infty dz\,   \text{Im}\left( W(z) M(z)\right) 
$~\footnote{It is useful to note  that $
\text{Im} W(z^\prime)=\om(z^\prime)/2$ and $\text{Re} W(z^\prime) = \frac{1}{2\pi} \int_0^\infty \om(z)  \left(  p.v. \frac{1}{z-z^\prime} + \frac{1}{z+z^\prime}  \right) \frac{z^2}{z^{\prime 2}} \, dz$.}  
and the   definition of the  functions $a_i(\eps)$, the Lagrangian in   \reef{lag1} simplifies into
\be
 L (\bM;  \Lambda) {=}2\gamma_3 -\lambda_2 c_2 - \lambda_3 c_3  - \lambda_4 c_4  + \int_0^\infty dz\, \frac{2}{\pi} \frac{c_2\lambda_3 {+} \lambda_4 c_3}{z^2}+ \int_0^\infty dz\,   \text{Im}\left( \widetilde{W}(z) M(z)\right) - \mu(z) U(z) 
\ee
where we have  defined
 \be
\widetilde{W} (z) \equiv W(z) + \frac{2}{\pi} \left(\frac{\lambda_2}{z^3}  - \frac{\lambda_4}{z^5}\right)
- i \frac{2}{\pi} \frac{\lambda_3}{z^4}
  \label{wt1}
 \, .
 \ee
 
 Now we are ready to find the extrema of the functional $L(\bM;\Lambda)\equiv \int dz {\cal L}(z)$. By using the  Euler-Lagrange e.o.m. $\partial_{M_c^*} {\cal L} = 0$, and find
 \be
M_c(z)= 2 i z  - \frac{ i z }{\mu(z)} \widetilde{W^*}(z)\, . 
\label{Mc1}\ee
Moreover, the Euler-Lagrange equation $\partial_{\gamma_3} {\cal L} = 0$ implies $\lambda_4=-1$, fixing one of the dual variables.
It is easy to check that $M_c$ is a minimum of $L(\bM;\Lambda)$.
Then, upon plugging the critical value of the amplitude $M_c$ back on the Lagrangian $L$ we are led to 
\be
d(\Lambda )  = \frac{1}{192}-\frac{ \lambda_2}{2}-\frac{\lambda_3}{16}   + \int_0^\infty  dz \Big[ \,  \frac{\lambda_3}{\pi z^2}-\frac{1}{8\pi z^2}  + 2z \Big( \text{Re}\widetilde{W}(z)  - \mu(z)   -  \frac{|\widetilde{W}(z)|^2}{4\mu(z)} \Big) \Big] \, ,  \label{d1}
\ee
where we have inserted the  LECs values  $\{c_2,c_3,c_4\}=\{\frac{1}{2},\, \frac{1}{16}, \,  1/192 - 2\gamma_3 \}$.

 The dual functional defined in \eqref{d1}, according to \reef{duality1},  gives a lower   bound on $\gamma_3$
for arbitrary  values of the dual variables  $\Lambda=\{\lambda_2,\lambda_3,\om(z),\mu(z)\} $.~\footnote{The functional in \reef{d1} it is only convergent for particular values of the multipliers. However, it is possible to ignore this subtlety working at $\epsilon>0$, using the definitions in \reef{arcs}, and taking the limit $\epsilon\to 0$ only at the end.} 
Next we will be able to find the maximal value of $d$ in \reef{d1} analytically. However, when considering more complicated problems in the sections below, it will be very useful to  perform a numerical search of the functions that maximise expressions like  \reef{d1}.

 \subsubsection{Analytic solution to Dual Problem I}
 \label{AnalyticDual}
We are now in a good position to solve  the  \hyperlink{pp2}{Dual Problem I } using the dual optimisation functional in \reef{d1}. We start by  finding the supremum  of \reef{d1} w.r.t.  varying $\mu(z)$ under the constraint $\mu(z)>0$. We  get  the critical function $\mu_c(z)=|\widetilde{W}(z)|/2$, which substituting back to \reef{d1} gives
\bea
D(W,\lambda_2,\lambda_3)\equiv \text{sup}_{\mu(z)} d(\Lambda)&=\frac{1}{192}-\frac{ \lambda_2}{2}-\frac{\lambda_3}{16} \nonumber \\
& + \int_0^\infty  dz \Big[  \frac{\lambda_3-\tfrac{1}{8}}{\pi} \frac{1}{z^2} +   2  z \Big( \text{Re}\widetilde{W}(z) - |\widetilde{W}(z)|   \Big) \Big] \, .  \label{D1}
\eea
Next we have to maximise the dual optimisation functional $D(W,\lambda_2,\lambda_3)$ over varying $W(z)$, $\lambda_2$ and $\lambda_3$.

Here it comes an interesting aspect of the dual functional for Wilson coefficients. 
The integrand in \reef{D1} has the following low energy expansion
\be
 \frac{\lambda_3-\tfrac{1}{8}}{\pi} \frac{1}{z^2} +     z \left( \text{Re}\widetilde{W}(z) - |\widetilde{W}(z)|   \right) =-\frac{(1-4  \lambda_3)^2}{8 \pi  z^2}+ O(z)\,. \label{pol2}
\ee
The factor $-(1-4  \lambda_3)^2\leq 0$ is negative for $\lambda_3 \in \mathds{R}$. 
Therefore upon integrating the latest expression we find that $D(W,\lambda_2,\lambda_3,\lambda_4)=-\infty$, unless the residue of the second order pole vanishes.
Thus, in order to maximize $D$ we must fix $\lambda_3=\tfrac{1}{4}$.
All in all, we get
\be
D(W,\lambda_2,1/4) =- \frac{1}{96} (1 +48\lambda_2)
  + \int_0^\infty  dz \left[ + \frac{1}{8\pi} \frac{1}{z^2} +  2   z \left( \text{Re}\widetilde{W}(z) - |\widetilde{W}(z)|   \right) \right]  \label{D2} \, , 
  \ee
which is a  nicely finite  dual  functional. We stress that  the finiteness of $D(W,\lambda_2,1/4)$, i.e. the "cancelation" of the value $-\infty$ by picking $\lambda_3=\tfrac{1}{4}$,  comes out naturally as a result of maximizing $D$ over varying $\lambda_i$'s.

To proceed further, we notice  that the maximum is attained by picking  $ \text{Im}W(z)=0$, which in turn using 
\reef{w1} implies  $ \text{Re}W(z)=0$.~\footnote{We can find the solution by varying $\text{Im}W(z)$ and $\text{Re}W(z)$ as independent field variables, and then check a posteriori that the solution falls inside the constraint \reef{w1}. }
 We are led to maximize the following functional over varying  $\lambda_2$ 
\be
D(0,\lambda_2,1/4) =  -\frac{\lambda_2}{2}-\frac{1}{96}
  + \frac{1}{\pi}\int_0^\infty  dz \left[  +\frac{1}{8} \frac{1}{z^2} +   \frac{4 \lambda_2}{ z^2}+\frac{4}{ z^4}  - \sqrt{  \left(\frac{4 \lambda_2}{ z^2}+\frac{4}{ z^4}\right)^2   +\frac{1}{  z^6} } \right]  \, . 
\ee
It is easy to check that $\lambda_2=-1/64$ is a local maximum of  $D(0,\lambda_2,1/4)$, and it is the unique zero of  $
f(\lambda_2) \equiv  \partial_{\lambda_2}D(0,\lambda_2,1/4)
 $ because  $f(\lambda_2)$
 is absolutely monotonic.~\footnote{This is expected: the dual problem is always concave for minimisation (convex for maximisation) independently of the properties of the primal. This follows from the definition of the Lagrangian and from the fact that point-wise extremization is a convexity-preserving operation.}  Therefore
\be
\text{sup}_{\{W,\lambda_2,\lambda_3\}} D(W,\lambda_2,\lambda_3) =  D(0,-1/64,1/4) =  -2 \frac{1}{768}  \, , 
\ee
in agreement with \reef{b1}!

We also find that the critical value of W is given by
 $
W_c(s)=\frac{2}{\pi}\frac{1}{s^5}  -\frac{1}{32\pi}\frac{1}{s^3}  -i  \frac{1}{2\pi}\frac{1}{s^4}  
 $. 
 Therefore using the fat that critical scattering amplitude \reef{Mc1} is given by  
\be
M_c(z)= 2 i z \left(1-  \widetilde{W_c^*}(z)/|\widetilde{W_c}(z)|\right) \,   \label{Mc2} 
\ee
we have
 $
M_c(s)= +\frac{4i s^2}{s+8i}
 $.~\footnote{This is similar to the Goldstino-like scattering amplitude introduced in \cite{Zamolodchikov:1991vx} -- similar bootstrap equations and bounds can be derived for the fermionic S-matrix $S(0)=-1$.}

The formulation presented in this section can be  generalised in order to bound the higher order LECs  $\gamma_5$ and $\gamma_7$ in \reef{lecs1}.
For these more involved dual problems, we also  find that the dual functional is finite when computed using the optimal $\lambda_i$'s, and the  extremal values of $\gamma_5$ and $\gamma_7$ coincide with  the primal optimisation problem  bounds of \cite{EliasMiro:2019kyf}.
Further details are given in appendix \ref{ggg}.

\section{Bounds on Flux Tubes}
\label{D4dual}

In $D \geq 4$ there are $D-2$ transverse directions to the flux-tube. This translates into $D-2$ Goldstone bosons that transform as vectors of  a $O(D-2)$ global symmetry. The scattering amplitude can be expressed in terms of three functions of the Mandelstam variable $s=(p_a+p_b)^2$
\beq
\mathbb{S}_{ab}^{dc}(s)=\sigma_1(s) \,  \delta_{ab}^{cd}+\sigma_2(s)  \, \delta_a^c \delta_b^d + \sigma_3(s) \,  \delta_a^d \delta_b^c =  \sigma_1(s)\dCinc{d}{b}{c}{a} + \, \sigma_2(s)\dCincDos{d}{b}{c}{a} +\, \sigma_3(s)\dCincTres{d}{b}{c}{a}
    \, . 
\eeq
These three functions describe  annihilation,   transmission and reflection of the vector index, as indicated by the diagrams. 
Crossing symmetry and real analyticity imply the following relations
\be
\sigma_1(-s^*)=\sigma_3(s)^*, \quad \sigma_2(-s^*)=\sigma_2(s)^*, \quad \sigma_3(-s^*)=\sigma_1(s)^*. \label{xreal}
\ee
Similarly to  the $D=3$ case, it is therefore possible to restrict the domain of these functions to the UHP without loss of generality.
The underlying $O(D-2)$ symmetry implies that  the  two-to-two S-matrix is diagonal when scattering two vectors in the  irreps.  of $O(D-2)$. Thus, the  suitable linear combinations 
\beq
S_\text{sing}=(D-2)\sigma_1+\sigma_2+\sigma_3,\quad S_\text{anti}=\sigma_2-\sigma_3, \quad S_\text{sym}=\sigma_2+\sigma_3,  \label{unitbase}
\eeq
 satisfy the diagonal unitary equation   
\be
|S_I(s)| \leq 1 \, ,\quad \text{for}  \quad  s\in(0,\infty) \, , 
\ee  where $I=\{\text{sing},\text{anti},\text{sym} \}$, and henceforth we will use capital index $I$ to denote these   channels.
The amplitudes, i.e. the interacting part of the S-matrix, is defined as usual $M_I=2 i s  \left(1- S_I\right)$. 

For our current purposes it is useful to introduce  a different basis:
\bea
M_1&= \frac{2M_\text{sing}+(D-4)M_\text{sym}  -(D-2)M_\text{anti} }{  4(D-2)}   \,   ,
\quad M_2=\frac{1}{2}\left( M_\text{sym}+M_\text{anti}\right) \, , \nonumber\\
\quad M_3&=\frac{2M_\text{sing}-DM_\text{sym}  +(D-2)M_\text{anti} }{  4i(D-2)}   \, ,   \label{xsymm1}
\eea
where crossing symmetry and real analyticity \reef{xreal} acts on the vector $(M_1,M_2,M_3)$ diagonally: $M_i(-s^*)=M_i(s)^*$.
In contrast to what happens in the single flavour case ($D=3$), unitarity does not act in a simple way in the basis where crossing-symmetry is diagonal. 

The low energy expansion of the flux tube (FT) amplitude defined in terms of the crossing symmetric components 
 reads
\bea
M_1^\text{FT}&=0\times s^2   \hspace{-2.4cm}&+&\, 0\times i s^3 \hspace{-2.4cm}&-&\, 2\beta_3\,  s^4+O(s^5)  \, , \nonumber\\
M_2^\text{FT}&=\frac{1}{2}\, s^2  \hspace{-2.4cm}&+&\, \frac{i}{16}\, s^3   \hspace{-2.4cm}&-&\left(\frac{1}{192}-2\alpha_3-2\beta_3\right)s^4+O(s^5)  \, , \nonumber\\
M_3^\text{FT}&=0\times s^2  \hspace{-2.4cm}&+&\, 2 i \alpha_2 \, s^3   \hspace{-2.4cm}&-&\, \frac{\alpha_2}{2} \, s^4+O(s^5) \, . 
\label{lowenergyDdim}
\eea
The coefficient $\alpha_2=\frac{D-26}{384\pi}$ is universal, depending only on the target space-time dimension.
The Wilson coefficients $\alpha_3$ and $\beta_3$ are related to the first two non-universal corrections to the $D=4$ flux tube action.

\subsection{The dual problem with flavor}

In this section we  apply the dual formalism to determine what is the allowed region in the $\{\alpha_3,\beta_3\}$ space excluding all the values of the Wilson coefficients that violate crossing, analyticity and unitarity.

In analogy to what we have done in Sec.~\ref{analit}, we express each coefficient of the low energy expansion of the amplitude $M_i=c_2^{(i)}s^2+\dots$ in terms of arc variables   of the respective amplitudes
\bea
a_2^{(i)}(\eps)&=\frac{2}{\pi}\int_\eps^\infty \frac{\im M_i(z)}{z^3}dz \, , \nonumber\\
a_3^{(i)}(\eps)&=-\frac{2}{\pi}\int_\eps^\infty \frac{\re M_i(z)-c_2^{(i)} z^2}{z^4}dz \, ,  \nonumber\\
a_4^{(i)}(\eps)&= - \frac{2}{\pi}\int_\eps^\infty \frac{\im M_i(z)-c_3^{(i)} z^3}{z^5}dz \, . \label{ontop}
\eea
Similarly to the previous section, the $c_n^{(i)}$ are read from the low energy expansion of   \reef{lowenergyDdim},
$M^\text{FT}_i=-c_2^{(i)} (-is)^2+c_3^{(i)}(-is)^3- c_4^{(i)}( -is)^4 + \dots$.
The notation will look slightly more Baroque because we need to carry with us the upper flavour index. Nevertheless the logic we follow is the same as in the $D=3$.

To find the boundaries of the $\{\alpha_3,\beta_3\}$ space we choose to minimize $\alpha_3$ 
at fixed $\beta_3$.~\footnote{
It is also possible to bound  a linear combination of the two Wilson coefficients $(\alpha_3,\beta_3)=(r\cos\theta,r\sin\theta)$, with $\theta$ fixed   and maximize the radius, similar to the \emph{radial optimization} of  \cite{Cordova:2019lot,Guerrieri:2020kcs}.}
Thus, we  formulate  the following (primal) problem in terms of dispersion relations:
\begin{subequations}
\begin{empheq}[box=\widefbox]{align}
\nonumber\\[-.1cm]
&  \text{\hypertarget{pp3}{Primal Problem II}:}  \nonumber\\[.1cm]
& \text{Minimize }   \alpha_3 \text{ varying } M_i(s)\text{  constrained by}   \nonumber \\
 &  \circ \,    U_I(s)   \equiv     2\, \text{Im} M_I(s) - \frac{1}{2z} |M_I(s)|^2  \geq 0, \  \text{ for } I\in \text{irreps, and } s>0 \,  ,  \label{unitd4}  \\
  &   \circ \,  
 \text{Disp}_i(s)  \equiv  \frac{1}{2}\text{Re}M_i(s)  - \frac{1}{2\pi  }  \int_0^\infty  \frac{s^2}{z^2}\, \text{Im}M_i(z) \left(p.v.\frac{1}{z-s} + \frac{1}{s+z} \right) dz =0 \quad  \nonumber \\ 
 & \phantom{\circ \, } \text{ for } i=1,2,3 \, \text{ and } \ s>0 \, ,  \label{anald4} \\[.1cm]
&  \circ \,   a_2^{(1)}(0)=0, \quad  \, a_3^{(1)}(0)=0,\quad  \ \, a_4^{(1)}(0)=2 \beta_3,  \label{lecd4} \\
&  \circ \,     a_2^{(2)}(0)=\frac{1}{2}, \quad  \hspace{-.01cm}  a_3^{(2)}(0)=\frac{1}{16}   ,  \ \,  \, \, a_4^{(2)}(0)= 1/192-2\alpha_3-2\beta_3, \label{lecd44} \\
&  \circ \,   a_2^{(3)}(0)=0,\quad  \, a_3^{(3)}(0)=2\alpha_2, \  \, \,  a_4^{(3)}(0)=\alpha_2/2  \, .  \label{primal4d}
\\[-.2cm] \nonumber
\end{empheq}
\end{subequations}
We remark that \reef{unitd4} is in the unitary basis \reef{unitbase}, while  \reef{anald4}  is in the crossing-symmetric basis \reef{xsymm1}.
In \reef{anald4} we took a twice subtracted dispersion relation for the three crossing-symmetric amplitudes.

The formulation of \hyperlink{pp3}{Primal Problem II} is in a nice form ready for dualization. 
Following the same strategy explained in Sec.~\ref{analit} we introduce a new Lagrangian
\be
L(\bM ;  \Lambda )  = 
\underbrace{\vphantom{\int_0^\infty }  2 \alpha_3}_\text{opt. goal} +
\underbrace{\vphantom{\int_0^\infty } \lambda^{(i)}_n (a_n^{(i)}(0)- c^{(i)}_n) }_\text{LECs constraints (\ref{lecd4}-\ref{primal4d})} +
\underbrace{ \int_0^\infty  \left[\, \om_i(z) \text{Disp}_i(z)  +  \mu_I(z) U_I(z)\,  \right]dz}_\text{analyticity and unitarity constraints} \, , \label{lag3}
\ee
with $I$ summed over $I\in  \{\text{sing},\text{anti},\text{sym} \}$,  $i$ over $i\in\{1,2,3\}$ in  the basis  of \reef{xsymm1}, and $n\in\{2,3,4\}$. 
The functions $\mu_I(z)\geq 0$ are non-negative, $\om_i(z)$'s are real   and   we have introduced  eight real dual variables $\lambda_n^{(i)}$, one for each of the eight low energy constraints in  (\ref{lecd4}-\ref{primal4d}). The primal and dual variables are collectively denoted by 
\be
\bM =\{\text{Re}M_i(z),\text{Im}M_i(z),\alpha_3 \}  \quad \text{and}  \quad \Lambda =\{\lambda_n^{(i)},\om_i(z),\mu_I(z)\} \, ,
\ee
respectively. 
It is useful to introduce  three analytic and anti-crossing symmetric functions $W_i(z)$ like \reef{w1}, 
 such that
$
\int_0^\infty \om_i(z)\text{Disp}_i(z)dz= $ $ \int_0^\infty \im (W_i(z)M_i(z))\,dz
$. 
It is also convenient to  further simplify the Lagrangian by defining  
 $
\widetilde{W}_i (z)  \pi/2\equiv  W_i(z) \pi/2+ \lambda_2^{(i)}/z^3- i  \lambda_3^{(i)}/z^4- \lambda_4^{(i)}/z^5$ with $i=1,2,3$, in order to absorb in $\widetilde{W}_i$ the contributions coming from the archs $a_n^{(i)}$'s. Then, we have
\be
L(\bM; \Lambda)=2 \alpha_3-\lambda_n^{(i)}c_n^{(i)}+\int_0^\infty  \frac{2}{\pi} \frac{\lambda_3^{(i)}c_2^{(i)}+\lambda_4^{(i)}c_3^{(i)}}{z^2} +\im (\widetilde W_i(z)M_i(z))+ \mu_I(z) U_I(z)dz\, ,  \label{lag111}
\ee
where we left implicit the sum over $I$, $i$ and $n$.~\footnote{E.g. $\lambda_n^{(i)}c_n^{(i)}= 1/96+2\alpha_3+2 \beta_3+\lambda_2^{(2)}/2-11\lambda_3^{(3)}/(96\pi)+2\beta_3 \lambda_4^{(1)}$.}
We introduce the dual functional
\be
d( \Lambda ) \equiv  \underset{\bM}{\phantom{p} \text{inf}\phantom{p}} L(\bM;\Lambda) \, .
\ee

Following analogous steps to the previous section and using equation \reef{duality1},
 it follows that 
\be
d( \Lambda )\leq  2  \alpha_3^* \, ,  \label{fineq}
\ee
where $ \alpha_3^*$ is the solution to \hyperlink{pp3}{Primal Problem II}.
The last equation provides the basis for  formulating
\vspace{-.3cm}
\begin{subequations}
\begin{empheq}[box=\widefbox]{align}
\nonumber\\[-.1cm]
&  \text{\hypertarget{pp4}{Dual Problem II}:}  \nonumber\\[.1cm]
& \text{Maximize} \  d(\Lambda)\text{ varying }  \Lambda=\{\om_i(z), \mu_I(z), \lambda_a^{(i)} \} \, ,   \text{constrained by } 
\mu_I(z)  > 0  \, . 
\end{empheq}
\end{subequations}
At this point it is simple to minimize over the primal variables $M_i$  and $\alpha_3$, and derive an analytical expression for the dual functional $d(\Lambda)$.
In particular, the equation of motion for $\alpha_3$ implies $\lambda_4^{(2)}=-1$.
The equations of motion for $M_I$ are derived in a similar way to the previous section. 

Given the simplicity of the dual objective, we can also maximize analytically over the multipliers $\mu_I \geq 0$.
After a bit of algebra we are lead to the following dual functional
\be
\text{sup}_{\mu_I(z)} d(\Lambda)
 =  
 -\lambda_n^{(i)}c_n^{(i)}+\int_0^\infty    \frac{2}{\pi} \frac{\lambda_3^{(i)}c_2^{(i)}+\lambda_4^{(i)}c_3^{(i)}}{z^2}  dz +\int_0^\infty  \frac{z}{2} \,  \Omega(z) dz \bigg|_{\lambda_4^{(2)}=-1} \, ,  \label{simpli}
 \ee
where  $\Omega(z)\equiv  4 \re \widetilde W_2- |\widetilde W_1{-}2\widetilde W_2{+}i \widetilde W_3|-\frac{2}{d-2} |\widetilde W_1{-}i \widetilde W_3| - \frac{1}{d-2} |(d{-}4)\widetilde W_1{+}2(d{-}2)\widetilde W_2{+}i d\widetilde W_3|  $. 

We want to emphasise that the  dual functional can be further maximized analytically by maximizing the residues of the poles of  the integrand in \reef{simpli}.
When the residues of the higher order poles in the expansion of $\tfrac{z}{2}\Omega(z)$ do not vanish, the dual functional is divergent with a definite sign, namely $d(\Lambda)\to -\infty$, hence 
providing a trivial (yet consistent) bound. Therefore, maximizing the residues turn out to be equivalent to set those to zero.
Explicitly, for $D=4$, 
\be
  \frac{z}{2} \Omega(z)=\frac{4-2\sqrt{ (\lambda_4^{(3)})^2+ 1}  -\sqrt{ (\lambda_4^{(3)})^2+ (\lambda_4^{(1)})^2}   
  - \sqrt{4+ (\lambda_4^{(1)})^2 +  (\lambda_4^{(3)})^2 +4 \lambda_4^{(1)}}    }{\pi  z^4}+ O(z^{-3}) \, . 
  \label{pole4}
\ee
In order to maximize  the residue in \reef{pole4}, we find  the critical values $\lambda_4^{(3)}=0$ and  $-2\leq \lambda_4^{(1)}\leq0$. 
For this choice of the dual variables, the coefficient of the $z^{-4}$ and $z^{-3}$ pole of   $\frac{z}{2} \Omega(z)$ vanishes.
Next we look for the $1/z^2$ and $1/z$ poles of the integrand in \reef{simpli} and cancel the corresponding residues  by maximising over 
$\lambda_a^{(i)}$. 
Solving the system of two equations for $\lambda_3^{(1)}$  and $\lambda_3^{(2)}$ and taking the real solution we find $(\lambda_3^{(1)},\, \lambda_3^{(2)})=(-\lambda_4^{(1)} /4,  \, +1/4)$
All in all we find that the values
\be
( \lambda_4^{(2)} , \,\lambda_4^{(3)} , \,\lambda_3^{(1)},\, \lambda_3^{(2)}  )=
(-1,0, -\lambda_4^{(1)} /4,  \, +1/4 )  \   \text{ and }  \ -2\leq \lambda_4^{(1)}\leq 0  \, ,  \label{minlamb}
\ee
maximize the dual functional, and  lead to a regular  integrand  in \reef{simpli} for $z\rightarrow 0$. 
The value $\lambda_4^{(3)}=0$ trivialises the constraint  
$a_4^{(3)}=\alpha_2/2$, which is fine because such constraint follows from unitarity (which we have already accounted for when integrating out $\mu_I(z)$'s in \reef{simpli}) once $a_3^{(i)}=c_3^{(i)}$ is satisfied.

Evaluating \reef{simpli} with the critical values in \reef{minlamb} we find
\be
 D(W_i,\lambda_2^{(i)},\lambda_3^{(3)},\lambda_4^{(1)}) 
 \equiv - \frac{1}{96} - \frac{\lambda_2^{(2)}}{2}+2 \alpha_2\lambda_3^{(3)}-2  \beta_3( \lambda_4^{(1)}+1)  +\int_0^\infty   \Big( \frac{1}{8 \pi  z^2}+ \frac{z}{2} \,  \Omega(z)  \Big) \,  dz  \, ,   \label{use1}
 \ee
for  $D=4$. 
All in all we are left with the
\vspace{-.3cm}
\begin{subequations}
\begin{empheq}[box=\widefbox]{align}
\nonumber\\[-.1cm]
&  \text{\hypertarget{pp5}{Simplified Dual Problem II}:}  \nonumber\\[.1cm]
& \text{Maximize} \   D(W_i,\lambda_2^{(i)},\lambda_3^{(3)},\lambda_4^{(1)})  \text{ varying }  \{W_i(z),  \lambda_2^{(i)} ,\, \lambda_3^{(3)}, \lambda_4^{(1)} \}   \, .  \hspace{1.5cm}  
\end{empheq}
\end{subequations}
We solve this problem in the next section.

\subsection{Bounds}
\label{secbounds}

According to \reef{fineq}, evaluating
$D(W_i , \lambda_a^{(i)})$  in  \reef{use1}  with arbitrary values of the dual variables, provides a rigorous bound to the minimal value of $2\alpha_3$ that can be achieved in \hyperlink{pp3}{Primal Problem II}.

In order to generate   bounds  that are close to optimality, we consider the following class of ansatzes
\be
W_j^\text{ans}(s)=  \frac{i}{s^2} \sum_{n=0}^{N_*} a^{(j)}_n \chi^n(s) \quad \text{where}\quad \chi(s)=\frac{i s_0-s}{i s_0+s} \, ,  \label{ansa}
\ee
for $j=1,2,3$,
and minimize   $D(W^\text{ans}_i,\lambda_2^{(i)},\lambda_3^{(3)},\lambda_4^{(1)})$   varying  $\{a_n^{(j)},  \lambda_a^{(i)}  \}$. 
The parameter $s_0$ is arbitrary, and  we  set $s_0=4$, in units of $\ell_s$.
We note that as $N_*\rightarrow \infty$, \reef{ansa} characterises an arbitrary anti-crossing symmetric function  $W^\text{ans}_j(s^*)=-W_j^\text{ans}(-s^*)$, analytic in the UHP of $s$, and that decays as $1/s^2$ as $s\rightarrow \infty $. Integrability at infinity of the dual function $D(W_i , \lambda_a^{(i)})$ requires an ansatz decaying as $1/s^3$, which  we achieve imposing additional linear constraints on the $a^{(j)}_n$'s. Imposing $\sum_n^{N_*}(-1)^n a_n^{(j)}=0$ guarantees $W_j^\text{ans}\sim 1/s^3$ as $s\rightarrow \infty$.  
We allow the ansatz to have  additional poles at threshold $s=0$, which are  allowed from general principles and the integrability of  \reef{use1}. Intuitively, the double pole we add is  `dual' to the double zero we find in the physical amplitude $M_i\sim s^2$.

\begin{figure}[t]
\centering
\includegraphics[width=0.8\textwidth]{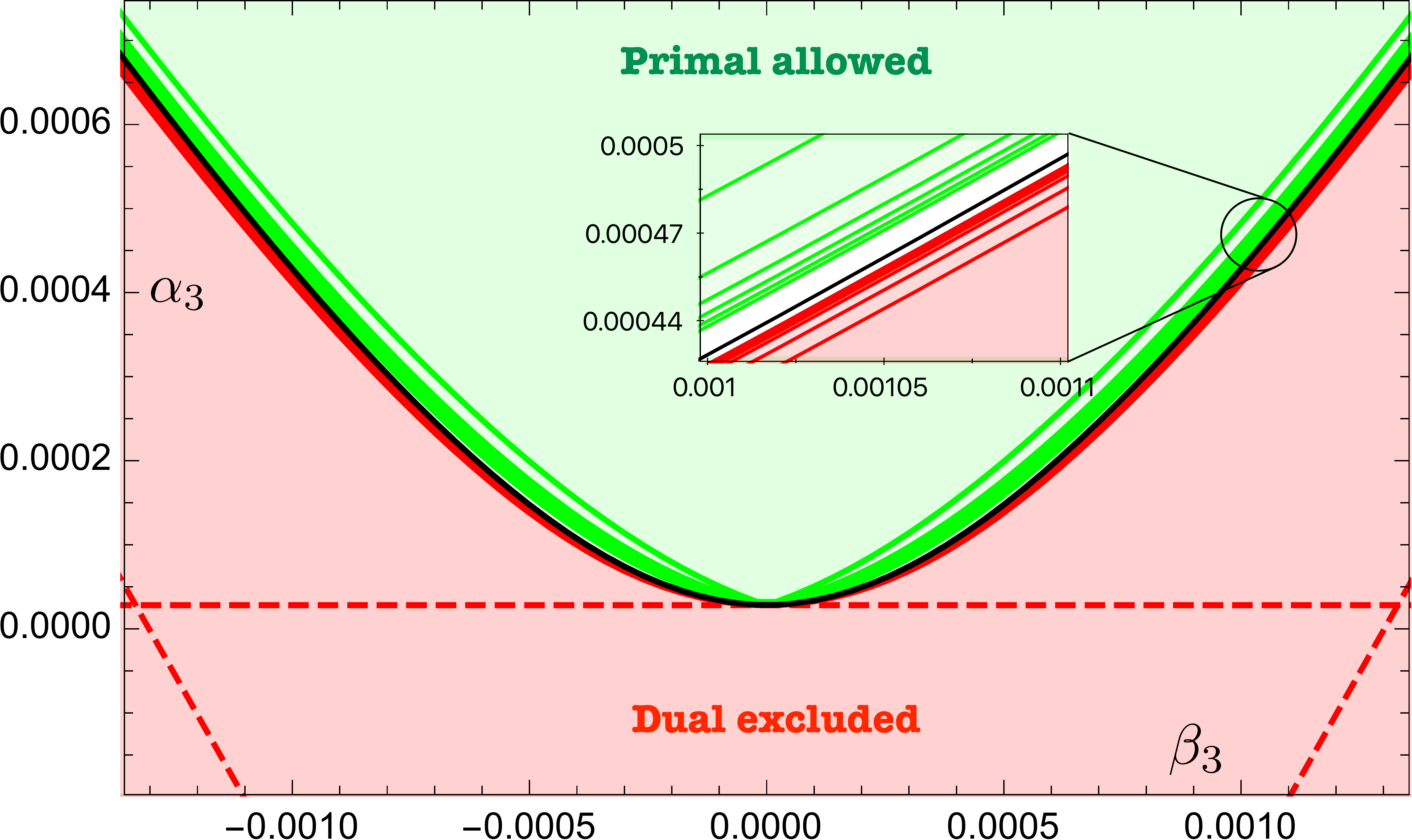}
\caption{Primal and dual bounds on the Wilson coefficients $\{\alpha_3, \beta_3\}$. The green region is allowed by primal numerics, the red region is excluded by the dual problem. The red lines are obtained solving the dual problem at fixed $\beta_3$ maximizing the dual functional for $N_*=5,10,\dots,30$; the dashed red lines are the analytic bounds obtained in \cite{EliasMiro:2019kyf}. The green lines denote the boundary at some fixed $N_\text{max}$ from $N_\text{max}=20,40,\dots, 120$; the black line is the power law extrapolation of primal numerics at $N_*\to\infty$. In the inset we zoom around a point of the boundary to appreciate better the convergence rate of dual numerics compared to the primal one.}
\label{figbound}
\end{figure}

The results of the dual minimisation  problem are shown in fig. \ref{figbound}. The different red lines correspond to values of $N^*=5,10,\dots,30$, and the region below, shaded in red colour, are the values of $\{\alpha_3,\beta_3\}$  that are rigorously excluded. 
Needless to say, $N_*=30$ signifies our best exclusion bound. 
Convergence is so fast that on the scale of the plot the red lines are all squeezed together. We have tried variational improvement with more sophisticated ansatzes~\footnote{Like for instance $
W_j^\text{ans}(s)=  \left( \frac{1}{(s+i z_j)^3}+\frac{1}{(s+i z_j)^2}\frac{R_j}{s}+\frac{1}{(s+i z_j)}\frac{R_{j+1}}{s^2} \right) \sum_{n=0}^{N_*} a^{(j)}_n \chi^n(s)$.} which   show a faster convergence. However,  for the maximal $N_*$ that we are reporting the difference between these variational improvements  is insignificant.

The green region results from primal numerics as in \cite{EliasMiro:2019kyf}. It is 
determined constructing primal solutions, namely minimising $\alpha_3$ at fixed $\beta_3$ in the space of amplitudes parametrized as in \reef{ans1} for different $n_\text{max}$ (the number of free parameters in the power series ansatz). In fig. \ref{figbound} the green lines correspond to values of $n_\text{max}=20,40,\dots,120$. 

Between the green and red lines there is a white space, see the zoomed in inset.
That is the duality gap which we do expect to vanish once optimality is attained (or when $n_\text{max}\rightarrow \infty$ and $ N^*\rightarrow \infty$).
We have also performed an extrapolation of the primal numerics in $n_\text{max}$~\footnote{Done with a simple-minded power-like fit $f(x)=a+b/x^c$, with three free parameters $\{a,b,c\}$.}, 
shown with a  black curve in fig. \ref{figbound}.
Interestingly, we find that the extrapolation of the primal falls nearly on top of the boundary of the exclusion region.

\section{Critical amplitudes and phase-shifts}
\label{cps}

The critical amplitudes are  obtained by minimising \reef{lag111} w.r.t $M_i$ and $\alpha_3$, and subsequently   evaluating the $\mu_I$ dependence by maximising $d(\Lambda)$. 
The procedure, which  is analogous to the one for $D=3$ that led to \reef{Mc2}, is simplified by working in the basis \footnote{ Notice the basis \eqref{unit_basis} is equivalent to the unitarity basis used in \cite{Cordova:2019lot} that makes unitarity trivial. }
\be
\mathds{W}_1 \equiv  \widetilde W_1-i \widetilde W_3 \ , \quad 
\mathds{W}_2  \equiv  \widetilde W_1-2 \widetilde W_2+i \widetilde W_3 \ , \quad 
\mathds{W}_3  \equiv  (D-4)\widetilde W_1+2 (D-2)\widetilde W_2+i d \widetilde W_3 \, . 
\label{unit_basis}
\ee
From the critical $M_i$'s we construct the S-matrices in each irrep. and, after a bit of algebra we find: 
\be
 \left(S^{D}_\text{sing} ,\,  S^{D}_\text{anti},  \,  S^{D}_\text{sym} \right) = \left(\frac{\mathds{W}_1^*}{|\mathds{W}_1|} ,\,  -\frac{\mathds{W}_2^*}{|\mathds{W}_2|}, \, \frac{\mathds{W}_3^*}{|\mathds{W}_3|} \right)  \, , 
 \label{dualS}
\ee
were the super index $D$ stands for dual. 
Interestingly,  the dual bounds provide the dual functions that  saturate $2\rightarrow 2$ unitarity  $|S^D_I| = 1$. 
Note however that the $S^{D}_\text{I}$'s  do not satisfy analyticity for generic values of the dual variables: this is only achieved  when the duality gap closes.

\begin{figure}[t!]
\centering
\includegraphics[width=0.95\textwidth]{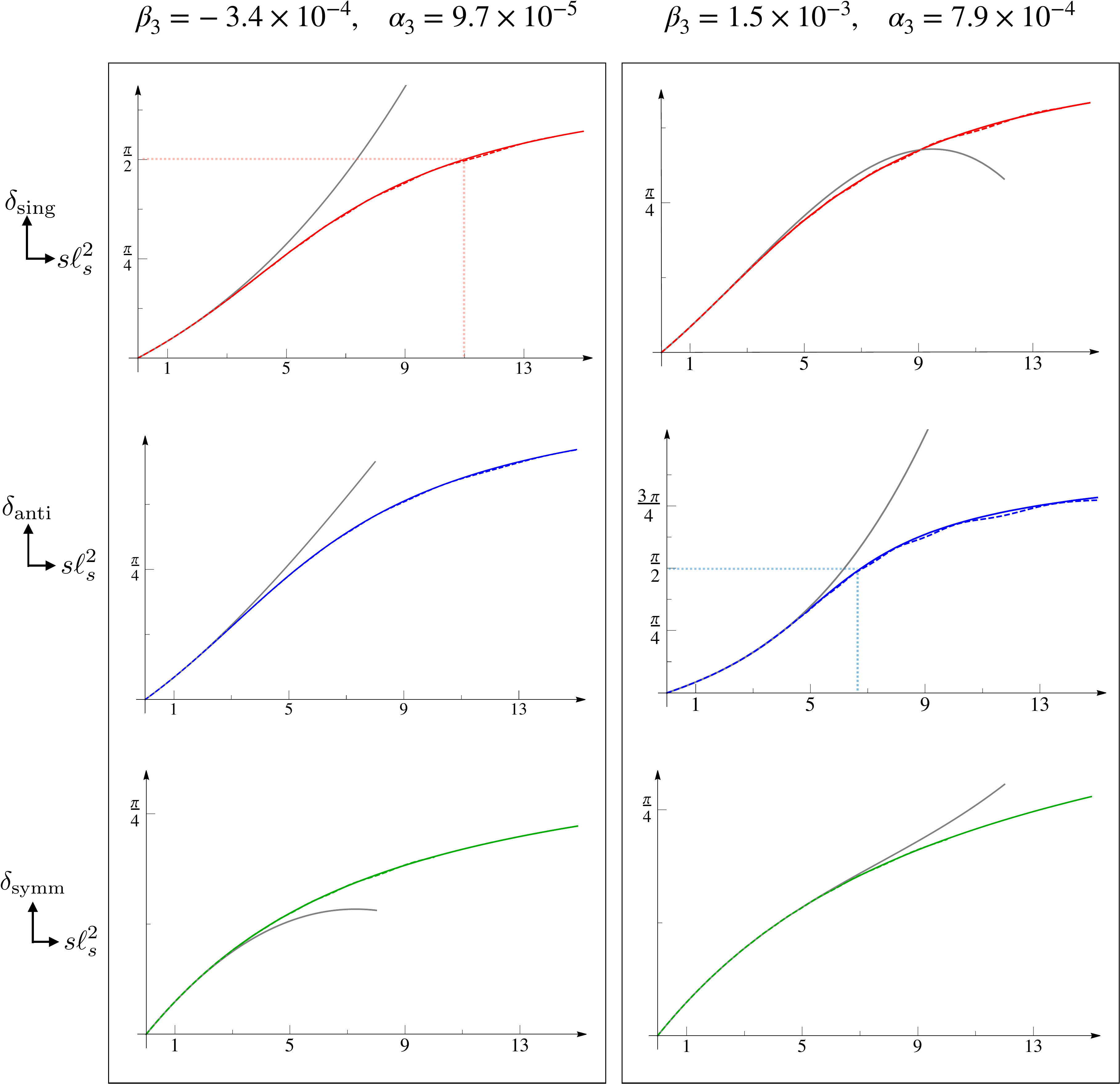}
\caption{ Phase shifts $\delta_I=\tfrac{1}{2i}\log{S_I}$ as a function of $s\ell_s^2$ for some irrep $I$, with $I$=\{singlet, antisymmetric, symmetric\} respectively in red, blue and green. In each plot, solid line is obtained from primal numerics with $N_\text{max}=120$, the dashed line is obtained from the dual with $N_*=30$. The gray lines are the predictions from the EFT up to two-loops.
The left panel shows the phase shifts for an arbitrary $\beta_3<0$: in the singlet channel there is a sharp resonance, signaled by the phase shifts passing through $\pi/2$. The right panel shows the phase shifts for a fixed $\beta_3>0$: in this case we see an axion resonance in the antisymmetric channel. Notice that for both points the EFT prediction agrees well with the non perturbative completion up to the scale set the by the lightest resonance, which, for this values of $\beta_3$ we chose appears dynamically around the naive cutoff scale $s^*=4/\ell_s^2$. 
}
\label{figps}
\end{figure}

In fig.~\ref{figps} we show the phase-shifts of the three-channels for two points in the boundary of fig.~\ref{figbound}.
In each plot we show three lines: the EFT  (gray), the dual (dashed) and primal (solid).
The dual S-matrix phases are obtained from  \reef{dualS} while  the  optimal primal 
phase-shifts are obtained following  \cite{EliasMiro:2019kyf}.
We find that the primal and dual S-matrix phases nicely coincidence.
We are showing a limited range of $s$ where the phases show the most interesting features. At larger $s$ the various phases eventually flatten. 

In the left panels we plot the phase shifts for a point along the boundary with $\beta_3<0$, in the right panels we do the same but for $\beta_3>0$. 
Those values of $\beta_3$ define two phases along the boundary of the allowed region in $\{\alpha_3,\beta_3\}$ separated by the integrable point at $\beta_3=0$~\cite{EliasMiro:2019kyf}.
The two phases differ by the presence of a sharp resonance respectively in the singlet (dilaton) and anti-symmetric channel (axion). 
In the $D=4$ case, these two phases are compatible with a symmetry of the crossing equations by exchanging singlet and anti-symmetric channels, which in turn exchanges the sign of $\beta_3$.
Interestingly, the axion branch agrees with the expectations from approximate integrability of the QCD flux-tube:
in \cite{EliasMiro:2019kyf} and in this work with the dual approach, we  find that the axion couples to the branons with the coupling dictated by the integrable theory \cite{Dubovsky:2015zey} that one would recover as the axions mass $m_a\rightarrow 0$.~\footnote{It is tempting to speculate that large $N$ Yang-Mills produces the integrable theory with $m_a\rightarrow 0$~\cite{Dubovsky:2015zey}. However, lattice MC simulaitons indicate that the axion mass achieves a positive value as $N\rightarrow \infty$~\cite{Athenodorou:2017cmw}. }

The plotted S-matrices allow analysing perturbative and non-perturbative physics. The perturbative physics amounts to the small momentum expansion  \reef{d4lecs}. 
Comparison of the EFT amplitude with the critical amplitude informs us of the cutoff.  We see that for the actual choice of $\{\alpha_3,\beta_3\}$ the EFT validity roughly coincides with the naive  EFT cutoff $s_*$ inferred in the IR  from $s_* \ell_s^2/4\approx 1$.
The actual cutoff is set dynamically by the non-perturbative phase-shifts shown in the singlet channel (left column) and antisymmetric channel (right).
These two abrupt phase-shifts signal the presence of an unstable resonance.

Finally we note that for $\beta_3=0$, we can find an analytic optimal solution of the dual problem. It is easy to check that
\beq
\lambda_4^{(1)}=-1,\quad \lambda_2^{(3)}=\alpha_2, \quad \lambda_2^{(1)}=\lambda_2^{(2)}=-\frac{1}{64}+16 \alpha_2^2,\quad \lambda_3^{(3)}=-8\alpha_2
\label{dulasolbeta30}
\eeq
with $W_i=0$ is a local maximum of the dual function $ D(W_i,\lambda_2^{(i)},\lambda_3^{(3)},\lambda_4^{(1)}) $, hence a global maximum because the dual functional is concave by definition.
The analytic value of the dual functional yields the exact inequality
\beq
\alpha_3\geq -\frac{1}{768}+4 \alpha_2^2.
\eeq
The $S$-matrix saturating this bound is explicitly integrable and can be obtained plugging the dual solution \eqref{dulasolbeta30} in the definition \eqref{dualS}
\beq
S^{D}_\text{sing}=S^{D}_\text{anti}=\frac{(32\alpha_2+i)s+8}{(32\alpha_2-i)s+8},\quad S^{D}_\text{sym}=\frac{(-32\alpha_2+i)s+8}{(-32\alpha_2-i)s+8}.
\eeq
This critical S-martrix nicely coincides with the one guessed in appendix C of \cite{EliasMiro:2019kyf}.

\section{Conclusions and outlook}
\label{cok}

In this work we have shown how to bound the space of two-dimensional EFTs through a
S-matrix  bootstrap approach. For concreteness we  have focused on the flux tube EFTs, which describe   the long effective string sector of three and four dimensional confining theories.

As  discussed in the introduction, positivity constraints on EFT Wilson coefficients has been a topic of intensive research for more than a decade. 
Due to the two dimensional nature of our system, we have been able to go beyond the positivity constraint  by considering the full two-particle sector unitarity equation \reef{unit2} instead of $\im M>0$. 
Nevertheless it is interesting to compare our methodologies with the positivity bounds widely employed in four dimensional EFTs. 
As a proof of concept we discuss the flux tube EFT for a single flavor (or $D=3$ flux-tube). 
The tree-level amplitude  is 
\be
M(s)= \texttt{c}_2 s^2 +  2 \gamma_3 s^4+ O(s^6) \, ,
\ee
where $\texttt{c}_2=1/2$ in the normalisation of the paper.
Therefore,  applying the widely-known EFT positivity dispersion relation~\cite{Adams:2006sv}~\footnote{While it is not essential to the logic low of our analysis, we remark that in two-dimensions there has been constructions of seemingly consistent UV complete  Lorentz invariant theories with the 'wrong sign' $\texttt{c}_2<0$~\cite{Dubovsky:2008bd,Cooper:2013ffa}, which exhibit superluminality.}, we  conclude
\be 
 \gamma_3>0 \, , \text{ at tree-level.}  \label{ntree}
\ee
In light of the perspective advocated in  \cite{Bellazzini:2020cot}, next we  improve the bound on $\gamma_3$ taking into account running effects, or loop corrections. 
For that purpose we define the  arc variables
\be
\text{arc}_2 =   \frac{2}{\pi} \int_\eps^\infty \frac{ \text{Im}M(z)}{z^3} dz = \texttt{c}_2+ O(\eps)  >0  \, , \quad 
\text{arc}_4 =  \frac{2}{\pi} \int_\eps^\infty \frac{ \text{Im} M(z)}{z^5} dz = \frac{\texttt{c}_2^2}{2\pi \eps}+\texttt{c}_4 + O(\eps)  >0 \, ,  \label{pos2}
\ee
where  the inequality signs follow from positivity  $\text{Im}M>0$,
and recall that  the loop-corrected amplitude  is given by
\be
M(s)= \texttt{c}_2s^2 + \underbrace{ \phantom{\Big|}i \texttt{c}_2^2 s^3/4  \phantom{\Big|}}_{\text{one-loop}} +\,  (2 \gamma_3\underbrace{-\frac{\texttt{c}_2^3}{24}}_{\text{two-loop}}) s^4+O(s^5) \, . \ \label{loopamp}
\ee 
The integrals in \reef{pos2} are done by deforming the contour as in fig.~\ref{contour},
 and $\texttt{c}_4$ is the coefficient of $s^4$ in  \reef{loopamp}.
Note that  due to peculiarities of two spacetime dimensions the massless cuts and naive $1/\pi$ loop factors are absent at this order (e.g. $s^3\log(s)+\text{crossing-symmetry}=s^3\log(s)-s^3\log(-s)=-i  \pi s^3$). 
Thus, after taking into account all loop corrections to the $O(s^4)$ amplitude, positivity of \reef{pos2} implies  
\be
\underbrace{2\gamma_3-\frac{\texttt{c}_2^3}{24}}_{\text{"running" } \gamma_3} > -\frac{\texttt{c}_2^2}{\eps \pi} + O(\eps) \, .  \label{pos3}
\ee
Two main points follow from \reef{pos3}: in the far IR $\eps{ \rightarrow} 0$ the constraint is satisfied due to IR~EFT unitarity $-\frac{\texttt{c}_2^2}{\eps \pi}<0$ (thus not sensitive to UV causality or analyticity constraints),  and  at intermediate energy scales 
the formula shows that loop corrections open a new region of parameter space allowing  $\gamma_3$ to be  negative. This is a sharp conclusion, which corrects the 
tree-level result \reef{ntree}.

Formula  \reef{pos3} does not allow us to  precisely  determine the value of the exact quantum bound on $\gamma_3$. Nevertheless,  we do expect that such bound must exist because  an  arbitrarily negative $\gamma_3$ would produce a negative phase \reef{lecs1}, which would signal non-analyticities in the UHP.~\footnote{Indeed,  analyticity in the UHP implies that the total integrated phase is non-negative $\int_{-\infty}^{+\infty} d\theta \partial_\theta 2\delta(\th)\geq0$ \cite{Doroud:2018szp}.}
As we have learned in this paper, such expectation  is precisely addressed by the  \emph{dual EFT bootstrap} approach 
 which sets the bound $\gamma_3 \geq -1/768$.
An amplitude with a $\gamma_3$ below such value is not feasible: it is either  non-analytic in the UHP or it violates unitarity for some energy regime.

The next key  step in the dual bootstrap  program is to generalise the approach developed in this work to higher dimensions. 
Recently in \cite{Guerrieri:2020bto,Guerrieri:2021ivu} it has been shown that the non perturbative bounds on pion-like and supergravity EFTs put strong constraints on the space of possible UV completions.
On the other hand, for those systems the precise determination of the feasible region in the space of Wilson coefficients using the numerical S-matrix Bootstrap is a  challenge.
It would be very interesting  to upgrade the dual EFT approach proposed in this work to higher dimensions and apply it to those and another phenomenologically relevant EFTs.

There are several questions the Dual Bootstrap might help to address in the context of two dimensional flux-tube EFTs.
In \cite{Dubovsky:2015zey} it was introduced  the so called Axionic String Ansatz (ASA) which 
proposes that   there are either no resonances  for the $D=3$ confining flux tube, or just the axion (the resonance in the antisymmetric channel) for the $D=4$ case. Positivity bounds for the $D=3$, under the ASA hypothesis, were derived already in \cite{EliasMiro:2019kyf}.~\footnote{See ref.~\cite{Conkey:2019blu} for a recent lattice calculation comparing the ASA for short strings against lattice MC simulations.}
For instance, in the $D=4$ case, we find that for  $\beta_3<0$ the optimal S-matrix contains a sharp dilaton resonance -- see fig. \ref{figps} -- and it would be excluded by incorporating the ASA into the Bootstrap constraints. We leave this exploration for the future.

Adding multi-particle processes to the bootstrap is a fascinating challenge both conceptually and numerically.
Two-dimensional flux-tube theories are simple enough yet rich of an interesting phenomenology that would justify the effort.
We believe the dual formulation might help tackling such a hard problem and perhaps single out the region where physical large-$N$ flux-tube theories might live.

We know that adding fermionic degrees of freedom and supersymmetry on the world-sheet of confining strings leads to a series of predictions for the low energy flux-tube dynamics and its S-matrix \cite{Cooper:2014noa}.
The scattering of supersymmetric gapped particles in two dimensions was studied in \cite{Bercini:2019vme} and the bound of allowed space of couplings showed interesting  geometric structures in that case.
It could happen that supersymmetric world-sheet theories lies at a special point in the  space of feasible Wilson coefficients. It would be interesting to study these theories with the dual bootstrap approach.

We have observed that the axion becomes lighter and that its coupling matches the integrable value as $\beta_3$ is increased along the boundary of fig.~\ref{figbound}. 
It is tempting to imagine that,  along this boundary,  the axion  mass $m_a$ decreases  following a technically natural trajectory which, within perturbation theory $s \ll \ell_s^{-1}$, could be defined as  the  integrable theory in~\cite{Dubovsky:2015zey} softly broken by the axion mass. 
It will be interesting to understand how generic is this feature by checking if the resonances observed in~\cite{EliasMiro:2019kyf}, and in this work, present an analogous pattern: the mass decreases along a section of the boundary of critical Wilson coefficients and the coupling to branons matches the integrable couplings of \cite{Dubovsky:2015zey}. 
As more couplings are turned on,  it would be interesting to explore the critical manifold of the dual EFT bootstrap. Are special points (cusps, edges, \dots) in this manifold of theories  close to the QCD string?, and what is the spectrum of resonances along such special trajectories?  
It will be fascinating to analyse these questions with the dual EFT bootstrap.

\subsection*{Acknowledgements}

We thank Sergei Dubovsky, Victor Gorbenko, Aditya Hebbar, Alexandre Homrich, Joao Penedones, Marco Serone, Amit Sever, Jacob Sonnenschein and Pedro Vieira 
for interesting discussions. We also thank Victor Gorbenko, Aditya Hebbar, Alexandre Homrich and Marco Serone for comments on the draft. 
AG is supported by The Israel Science Foundation (grant number 2289/18).

 
\appendix

\section{Numerical dual problem}
\label{ndp}

In this appendix we give more details about the numerical implementation of the dual problem focusing on the $D=4$ case.

As explained in sec.~\ref{D4dual}, the dual problem depend on a set of real variables $\vec \lambda$ and three anti-crossing holomorphic functions $W_i(s)$ in the UHP.
The space of $W_i(s)$ is infinite-dimensional, so we must truncate it choosing, for instance, a finite basis of functions.
A simple choice is the Taylor series expansion
\beq
W_i(s)=\frac{i}{s^2}\sum_{n=0}^{N_i^\prime} w_n^{(i)} \chi^n(s),
\eeq
where the function
\beq
\chi(s)=-\frac{s-i z_0}{s+i z_0}.
\eeq
maps the upper half plane to the unit disk with centre $i z_0$. \footnote{There is no obvious choice for $z_0$ a priori, though the rate of convergence of the numerical problem depend on its value.
For our numerics we have found empirically that $z_0 \sim 4$ gives the best convergence.}
 The prime means that we eliminate one constant $w_n^{(i)}$ in the sum to make $W_i \sim 1/s^3$ at large $s$. 
 This choice is dictated by the behaviour for $z\to \infty$ of the integrand in the dual functional definition \eqref{use1}.
The reader may also notice that the functional $W_i$ is not regular at $s=0$, but diverges as $W_i \sim i/s^2$. 
This divergence does not affect the convergence of the dual functional at the origin and it turns out that is needed to attain quickly the optimal bound. 

To compute the integral in \eqref{use1} numerically, we discretise the integrand on a grid of points using the Lagrange interpolation formula.
We first change variable mapping the positive energy axis $s>0$ to the segment $x \in [-1,1] $ using $s(x)=z_0 \tan(\tfrac{\pi}{4}(1+x))$, then we approximate the integrand
\beq
f(\vec w^{(i)},\vec \lambda| x)=\frac{z_0}{4}\frac{\pi}{\cos^2(\tfrac{\pi}{4}(1+x))}\left(-\frac{1}{8 \pi^2}\frac{1}{s(x)^2}+ \frac{s(x)}{2}\Omega(s(x))\right)
\eeq
by the interpolating polynomial of degree $N_\text{pts}$ passing through the $N_\text{pts}{+}1$ points $\{x_k\}$ \footnote{To run the  numerics we used $N_\text{pts}=300$.}
\beq
f(\vec w^{(i)},\vec \lambda| x) \approx \sum_{k=0}^{N_\text{pts}} f(\vec w^{(i)},\vec \lambda | x_k) \ell_k (x),
\label{fdef}
\eeq
where
\beq
\ell_k(x)=\prod_{m\neq k} \frac{x-x_m}{x_k-x_m}.
\eeq
For the interpolation points we use the set of Chebyschev nodes $x_k=\cos(\tfrac{k \pi}{N_\text{pts}+1})$.

Using \eqref{fdef} we obtain an approximated expression for the dual functional 
\beq
D(\vec w^{(i)},\vec \lambda_i) \approx- \frac{1}{96}-2\beta_3 - \frac{\lambda_2^{(2)}}{2}+2 \alpha_2\lambda_3^{(3)}-2  \beta_3 \lambda_4^{(1)}+ \sum_{k=0}^{N_\text{pts}} f(\vec w^{(i)},\vec \lambda | x_k) \int_{-1}^1 \ell_k (x)dx.
\label{Dapprox}
\eeq
To search for the maximum of $D$ we use the \verb Mathematica  built in function \verb FindMaximum. \

The discretised version of the dual objective in eq. \eqref{Dapprox}, used for the search provides a solution in terms of the dual variables $\{\vec w^{(i)},\vec \lambda\}$.
The numerical approximation does not affect the rigour of the bound since we can plug the solution found in the analytic expression \eqref{use1} obtaining a rigorous value.
We chose the number of points $N_\text{pts}$ large enough so that the difference between \eqref{Dapprox} and the analytic expression is much smaller compared to the typical values of the objective of our optimization.

\section{Analytic bounds on $\gamma_5$ and $\gamma_7$}
\label{ggg}

In this appendix we derive the analytic shape of the $D=3$ flux-tube ``Monolith" in \cite{EliasMiro:2019kyf}, namely the 3-dimensional allowed region in the $\gamma_{3,5,7}$ space using the 
dual technology developed in Sec. \ref{AnalyticDual}.

We start by considering the problem of minimizing $\gamma_5$ for any fixed value of $\gamma_3$, given the low energy expansion for the S-matrix
\beq
S(s)=e^{i\tfrac{s}{4}+i \gamma_3 s^3+i \gamma_5 s^5+i \gamma_7 s^7}+\mathcal{O}(s^8).
\label{SEFT}
\eeq
As explained in the main text, we fix the low energy ansatz using arcs sum rules \ref{arcs} for the amplitude $M=-2is(S-1)$, which explicitly yield 
\bea
a_{2k}(\epsilon)&=\frac{2(-1)^{k+1}}{\pi}\int_\epsilon^\infty\frac{\im M(z)-\sum_{m=3}^{2k-1} c_{m}\sin\tfrac{m\pi}{2}(-z)^{m}}{z^{2k+1}}dz,\nn\\
a_{2k+1}(\epsilon)&=\frac{2(-1)^{k}}{\pi}\int_\epsilon^\infty\frac{\re M(z)+\sum_{m=2}^{2k} c_{m} \cos\tfrac{m\pi}{2}(-z)^{m}}{z^{2(k+1)}}dz,
\eea
where the coefficients $c_m$ can be read off from the expansion
\beq
M(s)=-\sum_{m=2}^9(is)^{m} c_m(\gamma_i)+\mathcal{O}(s^9)
\eeq
using the definition in \eqref{SEFT}. 
Notice that these sum rules are valid if the amplitude we consider is analytic and crossing symmetric in the upper half plane.

So far, the derivation followed closely the one in Sec. \ref{AnalyticDual}. 
At this point we can take a shortcut.
We do not impose the dispersive constraint for any positive value of $s$, but we add just unitarity.
This is not a problem, of course, since a dual bound obtained imposing a subset of constraints is still a rigorous bound.
Nonetheless, it will not be generally optimal. 

The Lagrangian for this problem simply reads
\beq
L(M;\Lambda)=\gamma_5+\sum_{n=2}^6 \lambda_n (a_n(0)-c_n(\gamma_i))-\int_0^\infty \mu(z)U(z)dz,
\label{laggamma5}
\eeq
with $\mu\geq 0$, and
\beq
U(s)=2\im M(s)-\frac{|M(s)|^2}{2s}\geq 0.
\eeq
By solving the equations of motion we can solve for the $\re M$, $\im M$ and one of the $\lambda$'s
\bea
\frac{\delta L}{\delta \re M}=0 \implies& \re M=\frac{2}{\pi s^5 \mu}(s^2 \lambda_3-\lambda_5),\nn\\
\frac{\delta L}{\delta \im M}=0 \implies& \im M=\frac{2}{\pi s^6 \mu}(s^2 \lambda_4-s^4 \lambda_2+ \pi s^7 \mu-\tfrac{1}{2}),\nn\\
\frac{\partial L}{\partial \gamma_5}=0 \implies& \lambda_6 = \frac{1}{2}.
\label{LagEqGamma5}
\eea
Plugging this solution into the Lagrangian $L(M;\Lambda)$ yields the dual functional $d(\Lambda)$.
Before writing its explicit expression let us perform a further simplification. 

We recall that $d(\Lambda)$ is the objective of the dual problem which, in this example, provides lower bounds to the minimum value of $\gamma_5$ for any set of 
dual variables $\Lambda$.
However, due to the simplicity of the Lagrangian \eqref{laggamma5} we can also analytically maximise $d(\Lambda)$ wrt $\mu>0$, finding
\beq
\mu_c=\frac{1}{\pi s^3}\sqrt{\left(\lambda_2-\frac{\lambda_4}{s^2}+\frac{1}{2s^4} \right)^2+\left( \frac{\lambda_5}{s^3}-\frac{\lambda_3}{s}\right)^2}.
\eeq
Moreover, the function $D(\lambda_i)\equiv d(\lambda_i,\mu_c)$ is divergent for generic values of the dual variables. \footnote{It is reassuring to observe that $D(\lambda_i)=-\infty$ is still a lower bound, though a trivial one.}
We find that for this problem it is sufficient to fix $\lambda_5=-\tfrac{1}{8}$ to make sure that the dual functional $D(\lambda_i)$ converges, yielding explicitly
\bea
D(\lambda_i)&=\frac{1}{2^8 5!}-\frac{\gamma_3}{32}-\frac{\lambda_2}{2}-\frac{\lambda_3}{16}+\lambda_4\left(2\gamma_3-\frac{1}{192}\right)+\int_0^\infty \frac{dz}{\pi} \left(\frac{1}{16 z^4}+\frac{\lambda_3+\tfrac{1}{8}\lambda_4-\tfrac{1}{2^{10}}}{z^2}\right)\nn\\
&+\int_0^\infty dz\frac{4}{\pi z^2}\left( \frac{1}{2z^4}-\frac{\lambda_4}{z^2}+\lambda_2 - \sqrt{\left( \frac{1}{2z^4}-\frac{\lambda_4}{z^2}+\lambda_2\right)^2+ \left(\frac{1}{8 z^3}+\frac{\lambda_3}{z}\right)^2} \right).
\label{dualgamma5}
\eea
By numerical inspection it turns out that the maximum of $D(\lambda_i)$ is attained when the integrand in eq. \eqref{dualgamma5} vanishes.
Despite the non linearity of the integrand, it is possible to set it to zero choosing $\lambda_4=\tfrac{1}{128}-8 \lambda_3$ and $\lambda_2=32 \lambda_3^2$ leaving us with a function of $\lambda_3$ only
\beq
D(\lambda_3)=-16 \lambda_3^2-\left( \frac{1}{48}+16 \gamma_3 \right)-\frac{1}{122880}-\frac{\gamma_3}{64}.
\eeq
This is a concave function of $\lambda_3$ whose maximum is attained for $\lambda_3=-\tfrac{\gamma_3}{2}-\tfrac{1}{1536}$ producing the analytic inequality
\beq
D(\lambda_3)\leq 4 \gamma_3^2-\frac{\gamma_3}{192}-\frac{1}{737280} \leq \gamma_5.
\eeq

By definition, the local maximum we have found it is also global since 
the dual functional is a concave function of all the multipliers.

Once we find the optimal dual solution we can plug into the equation of motions \eqref{LagEqGamma5} and obtain the critical $S$-matrix 
\beq
S=1+\frac{i}{2s}M=\frac{8-32 \tilde \gamma_3 s^2+is}{8-32 \tilde \gamma_3 s^2-is},
\eeq
where $\tilde \gamma_3=\gamma_3+\tfrac{1}{768}$. For any fixed $\gamma_3$ this S-matrix is analytic in the upper half-plane and unitary with zeros whose location depend on the value of $\gamma_3$.
Hence, for this problem, we find that the dual optimal solution saturates all the constraints imposed and also the analyticity constraint we have not explicitly imposed.

The same argument can be applied to derive analytic bounds for the minimum $\gamma_7$ at fixed $\gamma_3$ and $\gamma_5$.
Here we just report the dual optimal solution
\bea
\lambda_2&=-\frac{(64 \tilde\gamma_5-256 \tilde\gamma_3^2+\tilde\gamma_3)^2}{524288\tilde\gamma_3^2}, \nn\\
\lambda_3&=-\frac{(\tilde\gamma_3(256\tilde\gamma_3-1)-64\tilde\gamma_5)(\tilde\gamma_3+64\tilde\gamma_5)}{32768 \tilde\gamma_3^2}, \nn\\
\lambda_4&=-\frac{3}{8192}-\frac{16\tilde\gamma_5^2+\tilde\gamma_3\tilde\gamma_5-2\tilde\gamma_3^3}{32\tilde\gamma_3^2}, \nn\\
\lambda_5&=\frac{1}{256}-\frac{\tilde \gamma_3}{2}+\frac{\tilde\gamma_5}{4\tilde \gamma_3}, \qquad \lambda_6=-\frac{3}{128}-\frac{\tilde\gamma_5}{\tilde\gamma_3}, \nn\\
\lambda_7&=\frac{1}{8}, \qquad \lambda_8=-\frac{1}{2}.
\eea

The bound on $\gamma_7$ is 
\beq
\gamma_7 \geq -\frac{1}{7340032}+\frac{\tilde\gamma_3}{4096}-\frac{\tilde\gamma_3^2}{16}+\frac{\tilde\gamma_5}{64}+\frac{\tilde\gamma_5^2}{\tilde\gamma_3}
\eeq
and the critical S-matrix
\beq
S(s)=\frac{((-8 + s) (-8 i + s) (8 + s) \tilde\gamma_3 - 256 s^3 \tilde\gamma_3^2 + 
 64 s^2 (-8 i + 
    s) \tilde\gamma_5)}{(-((-8 + s) (8 i + s) (8 + s) \tilde\gamma_3) + 
 256 s^3 \tilde\gamma_3^2 - 64 s^2 (8 i + s) \tilde\gamma_5)}.
\eeq

\section{Bonus:  critical manifold and log's}
\label{bonus}

The low energy expansion of the  $D=4$ flux tube S-matrix is analytic up to $O(s^5)$  \cite{Cooper:2014noa}. The  first non-analytic terms are of the form  $s^5\log s$, and are fixed by unitarity. 
At $O(s^4)$ there is a new non-universal parameter $\alpha_4$ ($O(s^5)$ in the M-matrix), and  $O(s^5)$ there are two  new non-universal coefficients $\{\alpha_5,\beta_5\}$ appearing in the $S$-matrix (hence $O(s^6)$ in the $M$-matrix). 
In this section we  extend the dual functional introduced in the main text  incorporating the parametrisation of the low energy expansion up to $O(s^6)$. 

It turns out that \reef{lowenergyDdim} generalises into
\bea
M_1^\text{FT}&=0\times s^2+0\times i s^3-2\beta_3 s^4 -\frac{1}{2} i \beta _3 s^5+\left(\frac{1}{16}\beta_3-2{\blu \beta_5}-\frac{4  }{\pi }\alpha_2 \beta_3 \log (-is)\right) s^6 +O(s^7)\\
M_2^\text{FT}&=\frac{1}{2}s^2+\frac{i}{16}s^3-\left(\frac{1}{192}-2\alpha_3-2\beta_3\right)s^4-\frac{i}{2}\left(\frac{1}{1536}- 2\alpha _2^2- \alpha _3- \beta _3 \right) s^5 \nonumber\\
&+ \left( \frac{1}{61440}+2 {\blu \alpha_5}+2{\blu \beta_5} -\frac{ (2\alpha_2)^2+\alpha_3+\beta_3}{16}-\frac{4 }{\pi }\alpha_2 \beta_3 \log (-i s)  \right) s^6+ O(s^7) \\
M_3^\text{FT}&=0\times s^2+2 i \alpha_2 s^3-\frac{\alpha_2}{2} s^4-  i  \left(\frac{ \alpha _2}{16}-2\,   {\roig \alpha_4}\right)   s^5 + \left(\frac{\alpha_2}{192}-2  \alpha_2 \alpha_3-\frac{{\roig \alpha_4}}{2}\right) s^6+O(s^7),
\label{lowenergyDdim2}
\eea 
where we are using the crossing-symmetric basis introduced in \reef{xsymm1}, and we have indicated in {\roig red} and {\blu blue} the appearance of the higher order non-universal parameters $\alpha_4$ and $\{\alpha_5,\beta_5\}$.

Once more, we repeat the steps to formulate the dual functional. We define the Lagrangian
\be
L(M_i ;  \Lambda )  = 
\underbrace{\vphantom{\int_0^\infty } o.g.}_\text{opt. goal} +
\underbrace{\vphantom{\int_0^\infty } \lambda^{(i)}_n (a_n^{(i)}(0)- c^{(i)}_n) }_\text{LECs constraints \reef{lowenergyDdim2}} +
\underbrace{ \int_0^\infty  \left[\, \om_i(z) \text{Disp}_i(z)  +  \mu_I(z) U_I(z)\,  \right]dz}_\text{analyticity and unitarity constraints} \, . \label{lagB}
\ee
where $\Lambda$ collectively denotes all the Lagrange multipliers $\{\lambda_n^{(i)},\om_i,\mu_I\}$;   the $c_n^{(i)}$ are read from the low energy expansion $M^\text{FT}_i = \sum_n^5 s^n c_n^{(i)}+c_6^{(i)}s^6+c_{6,1}^{(i)}s^6\log(-is)+O(s^7)$ in \reef{lowenergyDdim2}; and on top of \reef{ontop} we are using
\bea
a_5^{(i)}(\eps) &=  \frac{2}{\pi}\int_\eps^\infty \frac{\re M_i(z)-c_2^{(i)} z^2+c_4^{(i)}z^4}{z^6}dz \, ,  \\
a_6^{(i)}(\eps) &=  \frac{2}{\pi}\int_\eps^\infty \frac{\im M_i(z)-c_3^{(i)} z^3+c_5^{(i)}z^5-c_{6,1}^{(i)}z^6\pi/2}{z^7}dz  \, .
\eea


After going through the by now familiar algebra  we are led to  the following dual functional
\bea
D(\Lambda)\equiv \text{inf}_{\mu_I(z)} d(\Lambda)
 &= o.g. -\lambda_n^{(i)}c_n^{(i)} +\frac{2}{\pi}  \int_0^\infty dz\,  \frac{\lambda_3^{(i)}c_2^{(i)}+\lambda_4^{(i)}c_3^{(i)}+\lambda_5^{(i)}c_4^{(i)}  +\lambda_6^{(i)}c_5^{(i)}  }{z^2}   \, ,  \\
 & -   \frac{2}{\pi} \int_0^\infty dz\frac{\lambda_5^{(i)}c_2^{(i)}+  \lambda_6^{(i)}c_3^{(i)}  }{z^4}-\int_0^\infty dz \frac{\lambda_6^{(i)}c_{6,1}^{(i)}}{z}+\int_0^\infty dz   \frac{z}{2} \,  \Omega(z)  \label{lastD}
 \eea
where $\Omega(z)\equiv  4 \re \widetilde W_2+ |\widetilde W_1{-}2\widetilde W_2{+}i \widetilde W_3|+\frac{2}{d-2} |\widetilde W_1{-}i \widetilde W_3| + \frac{1}{d-2} |(d{-}4)\widetilde W_1{+}2(d{-}2)\widetilde W_2{+}i d\widetilde W_3|  $ and  we have defined   
 $
\widetilde{W}_i (z)  \pi/2\equiv  W_i(z) \pi/2+ \lambda_2^{(i)}/z^3- i  \lambda_3^{(i)}/z^4- \lambda_4^{(i)}/z^5+i\lambda_5^{(i)}/z^6+ \lambda_6^{(i)}/z^7$ with $i=1,2,3$.
By the same reasoning explained in the sections above, lower bounds on the minimal value of $o.g.$ can be placed by evaluating the dual functional \reef{lastD}, and the most stringent bound are found by maximising $D(\Lambda)$ over the Lagrange multipliers.

Our next task is to remove the potential singularities $D(\Lambda)\rightarrow -\infty$ by maximising over the $\lambda_n^{(i)}$'s.
Again we find that dual functional is nicely finite at the maxima. In particular by fixing
\be
(\lambda_6^{(2)},\lambda_6^{(3)}, \lambda_5^{(1)},\lambda_5^{(2)},\lambda_5^{(3)} ) = (\lambda_6^{(1)},0,-\lambda_6^{(1)}/4,-\lambda_6^{(1)}/4,4 \alpha_2\lambda_6^{(1)})
\ee
and $\lambda_6^{(1)}>0$  the integrand in \reef{lastD} is analytic around $z=0$.  
We have obtained bounds  -- taking $(o.g., \lambda_6^{(1)},\beta_3)=(2 \alpha_5,1,0)$ and  scanning over $\alpha_4$ -- but we leave for the future the detailed investigation of the critical manifold.


\small

\bibliography{biblio}

\providecommand{\href}[2]{#2}\begingroup\raggedright\begin{thebibliography}{10}

\bibitem{Adams:2006sv}
A.~Adams, N.~Arkani-Hamed, S.~Dubovsky, A.~Nicolis, and R.~Rattazzi,
  ``{Causality, analyticity and an IR obstruction to UV completion},''
  \href{http://dx.doi.org/10.1088/1126-6708/2006/10/014}{{\em JHEP} {\bfseries
  10} (2006) 014}, \href{http://arxiv.org/abs/hep-th/0602178}{{\ttfamily
  arXiv:hep-th/0602178}}.

\bibitem{Pham:1985cr}
T.~N. Pham and T.~N. Truong, ``{Evaluation of the Derivative Quartic Terms of
  the Meson Chiral Lagrangian From Forward Dispersion Relation},''
  \href{http://dx.doi.org/10.1103/PhysRevD.31.3027}{{\em Phys. Rev. D}
  {\bfseries 31} (1985) 3027}.

\bibitem{Pennington:1994kc}
M.~R. Pennington and J.~Portoles, ``{The Chiral Lagrangian parameters, l1, l2,
  are determined by the rho resonance},''
  \href{http://dx.doi.org/10.1016/0370-2693(94)01551-M}{{\em Phys. Lett. B}
  {\bfseries 344} (1995) 399--406},
  \href{http://arxiv.org/abs/hep-ph/9409426}{{\ttfamily arXiv:hep-ph/9409426}}.

\bibitem{Ananthanarayan:1994hf}
B.~Ananthanarayan, D.~Toublan, and G.~Wanders, ``{Consistency of the chiral
  pion pion scattering amplitudes with axiomatic constraints},''
  \href{http://dx.doi.org/10.1103/PhysRevD.51.1093}{{\em Phys. Rev. D}
  {\bfseries 51} (1995) 1093--1100},
  \href{http://arxiv.org/abs/hep-ph/9410302}{{\ttfamily arXiv:hep-ph/9410302}}.

\bibitem{Komargodski:2011vj}
Z.~Komargodski and A.~Schwimmer, ``{On Renormalization Group Flows in Four
  Dimensions},'' \href{http://dx.doi.org/10.1007/JHEP12(2011)099}{{\em JHEP}
  {\bfseries 12} (2011) 099}, \href{http://arxiv.org/abs/1107.3987}{{\ttfamily
  arXiv:1107.3987 [hep-th]}}.

\bibitem{Bellazzini:2016xrt}
B.~Bellazzini, ``{Softness and amplitudes\textquoteright{} positivity for
  spinning particles},'' \href{http://dx.doi.org/10.1007/JHEP02(2017)034}{{\em
  JHEP} {\bfseries 02} (2017) 034},
  \href{http://arxiv.org/abs/1605.06111}{{\ttfamily arXiv:1605.06111
  [hep-th]}}.

\bibitem{Cheung:2016yqr}
C.~Cheung and G.~N. Remmen, ``{Positive Signs in Massive Gravity},''
  \href{http://dx.doi.org/10.1007/JHEP04(2016)002}{{\em JHEP} {\bfseries 04}
  (2016) 002}, \href{http://arxiv.org/abs/1601.04068}{{\ttfamily
  arXiv:1601.04068 [hep-th]}}.

\bibitem{Luty:2012ww}
M.~A. Luty, J.~Polchinski, and R.~Rattazzi, ``{The $a$-theorem and the
  Asymptotics of 4D Quantum Field Theory},''
  \href{http://dx.doi.org/10.1007/JHEP01(2013)152}{{\em JHEP} {\bfseries 01}
  (2013) 152}, \href{http://arxiv.org/abs/1204.5221}{{\ttfamily arXiv:1204.5221
  [hep-th]}}.

\bibitem{Distler:2006if}
J.~Distler, B.~Grinstein, R.~A. Porto, and I.~Z. Rothstein, ``{Falsifying
  Models of New Physics via WW Scattering},''
  \href{http://dx.doi.org/10.1103/PhysRevLett.98.041601}{{\em Phys. Rev. Lett.}
  {\bfseries 98} (2007) 041601},
  \href{http://arxiv.org/abs/hep-ph/0604255}{{\ttfamily arXiv:hep-ph/0604255}}.

\bibitem{Englert:2019zmt}
C.~Englert, G.~F. Giudice, A.~Greljo, and M.~Mccullough, ``{The
  $\hat{H}$-Parameter: An Oblique Higgs View},''
  \href{http://dx.doi.org/10.1007/JHEP09(2019)041}{{\em JHEP} {\bfseries 09}
  (2019) 041}, \href{http://arxiv.org/abs/1903.07725}{{\ttfamily
  arXiv:1903.07725 [hep-ph]}}.

\bibitem{Bellazzini:2017fep}
B.~Bellazzini, F.~Riva, J.~Serra, and F.~Sgarlata, ``{Beyond Positivity Bounds
  and the Fate of Massive Gravity},''
  \href{http://dx.doi.org/10.1103/PhysRevLett.120.161101}{{\em Phys. Rev.
  Lett.} {\bfseries 120} no.~16, (2018) 161101},
  \href{http://arxiv.org/abs/1710.02539}{{\ttfamily arXiv:1710.02539
  [hep-th]}}.

\bibitem{Alberte:2020bdz}
L.~Alberte, C.~de~Rham, S.~Jaitly, and A.~J. Tolley, ``{QED positivity
  bounds},'' \href{http://arxiv.org/abs/2012.05798}{{\ttfamily arXiv:2012.05798
  [hep-th]}}.

\bibitem{Bellazzini:2019bzh}
B.~Bellazzini, F.~Riva, J.~Serra, and F.~Sgarlata, ``{Massive Higher Spins:
  Effective Theory and Consistency},''
  \href{http://dx.doi.org/10.1007/JHEP10(2019)189}{{\em JHEP} {\bfseries 10}
  (2019) 189}, \href{http://arxiv.org/abs/1903.08664}{{\ttfamily
  arXiv:1903.08664 [hep-th]}}.

\bibitem{Gu:2020ldn}
J.~Gu, L.-T. Wang, and C.~Zhang, ``{An unambiguous test of positivity at lepton
  colliders},'' \href{http://arxiv.org/abs/2011.03055}{{\ttfamily
  arXiv:2011.03055 [hep-ph]}}.

\bibitem{deRham:2018qqo}
C.~de~Rham, S.~Melville, A.~J. Tolley, and S.-Y. Zhou, ``{Positivity Bounds for
  Massive Spin-1 and Spin-2 Fields},''
  \href{http://dx.doi.org/10.1007/JHEP03(2019)182}{{\em JHEP} {\bfseries 03}
  (2019) 182}, \href{http://arxiv.org/abs/1804.10624}{{\ttfamily
  arXiv:1804.10624 [hep-th]}}.

\bibitem{Arkani-Hamed:2020blm}
N.~Arkani-Hamed, T.-C. Huang, and Y.-T. Huang, ``{The EFT-Hedron},''
  \href{http://arxiv.org/abs/2012.15849}{{\ttfamily arXiv:2012.15849
  [hep-th]}}.

\bibitem{Green:2019tpt}
M.~B. Green and C.~Wen, ``{Superstring amplitudes, unitarily, and Hankel
  determinants of multiple zeta values},''
  \href{http://dx.doi.org/10.1007/JHEP11(2019)079}{{\em JHEP} {\bfseries 11}
  (2019) 079}, \href{http://arxiv.org/abs/1908.08426}{{\ttfamily
  arXiv:1908.08426 [hep-th]}}.

\bibitem{Bellazzini:2020cot}
B.~Bellazzini, J.~Elias~Mir\'o, R.~Rattazzi, M.~Riembau, and F.~Riva,
  ``{Positive Moments for Scattering Amplitudes},''
  \href{http://arxiv.org/abs/2011.00037}{{\ttfamily arXiv:2011.00037
  [hep-th]}}.

\bibitem{Tolley:2020gtv}
A.~J. Tolley, Z.-Y. Wang, and S.-Y. Zhou, ``{New positivity bounds from full
  crossing symmetry},'' \href{http://arxiv.org/abs/2011.02400}{{\ttfamily
  arXiv:2011.02400 [hep-th]}}.

\bibitem{Caron-Huot:2020cmc}
S.~Caron-Huot and V.~Van~Duong, ``{Extremal Effective Field Theories},''
  \href{http://arxiv.org/abs/2011.02957}{{\ttfamily arXiv:2011.02957
  [hep-th]}}.

\bibitem{Caron-Huot:2021rmr}
S.~Caron-Huot, D.~Mazac, L.~Rastelli, and D.~Simmons-Duffin, ``{Sharp
  Boundaries for the Swampland},''
  \href{http://arxiv.org/abs/2102.08951}{{\ttfamily arXiv:2102.08951
  [hep-th]}}.

\bibitem{Bern:2021ppb}
Z.~Bern, D.~Kosmopoulos, and A.~Zhiboedov, ``{Gravitational Effective Field
  Theory Islands, Low-Spin Dominance, and the Four-Graviton Amplitude},''
  \href{http://arxiv.org/abs/2103.12728}{{\ttfamily arXiv:2103.12728
  [hep-th]}}.

\bibitem{Chiang:2021ziz}
L.-Y. Chiang, Y.-t. Huang, W.~Li, L.~Rodina, and H.-C. Weng, ``{Into the
  EFThedron and UV constraints from IR consistency},''
  \href{http://arxiv.org/abs/2105.02862}{{\ttfamily arXiv:2105.02862
  [hep-th]}}.

\bibitem{Paulos:2016fap}
M.~F. Paulos, J.~Penedones, J.~Toledo, B.~C. van Rees, and P.~Vieira, ``{The
  S-matrix bootstrap. Part I: QFT in AdS},''
  \href{http://dx.doi.org/10.1007/JHEP11(2017)133}{{\em JHEP} {\bfseries 11}
  (2017) 133}, \href{http://arxiv.org/abs/1607.06109}{{\ttfamily
  arXiv:1607.06109 [hep-th]}}.

\bibitem{Paulos:2016but}
M.~F. Paulos, J.~Penedones, J.~Toledo, B.~C. van Rees, and P.~Vieira, ``{The
  S-matrix bootstrap II: two dimensional amplitudes},''
  \href{http://dx.doi.org/10.1007/JHEP11(2017)143}{{\em JHEP} {\bfseries 11}
  (2017) 143}, \href{http://arxiv.org/abs/1607.06110}{{\ttfamily
  arXiv:1607.06110 [hep-th]}}.

\bibitem{Paulos:2017fhb}
M.~F. Paulos, J.~Penedones, J.~Toledo, B.~C. van Rees, and P.~Vieira, ``{The
  S-matrix bootstrap. Part III: higher dimensional amplitudes},''
  \href{http://dx.doi.org/10.1007/JHEP12(2019)040}{{\em JHEP} {\bfseries 12}
  (2019) 040}, \href{http://arxiv.org/abs/1708.06765}{{\ttfamily
  arXiv:1708.06765 [hep-th]}}.

\bibitem{Guerrieri:2018uew}
A.~L. Guerrieri, J.~Penedones, and P.~Vieira, ``{Bootstrapping QCD Using Pion
  Scattering Amplitudes},''
  \href{http://dx.doi.org/10.1103/PhysRevLett.122.241604}{{\em Phys. Rev.
  Lett.} {\bfseries 122} no.~24, (2019) 241604},
  \href{http://arxiv.org/abs/1810.12849}{{\ttfamily arXiv:1810.12849
  [hep-th]}}.

\bibitem{Homrich:2019cbt}
A.~Homrich, J.~a. Penedones, J.~Toledo, B.~C. van Rees, and P.~Vieira, ``{The
  S-matrix Bootstrap IV: Multiple Amplitudes},''
  \href{http://dx.doi.org/10.1007/JHEP11(2019)076}{{\em JHEP} {\bfseries 11}
  (2019) 076}, \href{http://arxiv.org/abs/1905.06905}{{\ttfamily
  arXiv:1905.06905 [hep-th]}}.

\bibitem{Karateev:2019ymz}
D.~Karateev, S.~Kuhn, and J.~a. Penedones, ``{Bootstrapping Massive Quantum
  Field Theories},'' \href{http://dx.doi.org/10.1007/JHEP07(2020)035}{{\em
  JHEP} {\bfseries 07} (2020) 035},
  \href{http://arxiv.org/abs/1912.08940}{{\ttfamily arXiv:1912.08940
  [hep-th]}}.

\bibitem{Hebbar:2020ukp}
A.~Hebbar, D.~Karateev, and J.~Penedones, ``{Spinning S-matrix Bootstrap in
  4d},'' \href{http://arxiv.org/abs/2011.11708}{{\ttfamily arXiv:2011.11708
  [hep-th]}}.

\bibitem{Correia:2020xtr}
M.~Correia, A.~Sever, and A.~Zhiboedov, ``{An Analytical Toolkit for the
  S-matrix Bootstrap},'' \href{http://arxiv.org/abs/2006.08221}{{\ttfamily
  arXiv:2006.08221 [hep-th]}}.

\bibitem{EliasMiro:2019kyf}
J.~Elias~Mir\'o, A.~L. Guerrieri, A.~Hebbar, J.~a. Penedones, and P.~Vieira,
  ``{Flux Tube S-matrix Bootstrap},''
  \href{http://dx.doi.org/10.1103/PhysRevLett.123.221602}{{\em Phys. Rev.
  Lett.} {\bfseries 123} no.~22, (2019) 221602},
  \href{http://arxiv.org/abs/1906.08098}{{\ttfamily arXiv:1906.08098
  [hep-th]}}.

\bibitem{Guerrieri:2020bto}
A.~Guerrieri, J.~Penedones, and P.~Vieira, ``{S-matrix Bootstrap for Effective
  Field Theories: Massless Pions},''
  \href{http://arxiv.org/abs/2011.02802}{{\ttfamily arXiv:2011.02802
  [hep-th]}}.

\bibitem{Bose:2020shm}
A.~Bose, P.~Haldar, A.~Sinha, P.~Sinha, and S.~S. Tiwari, ``{Relative entropy
  in scattering and the S-matrix bootstrap},''
  \href{http://dx.doi.org/10.21468/SciPostPhys.9.5.081}{{\em SciPost Phys.}
  {\bfseries 9} (2020) 081}, \href{http://arxiv.org/abs/2006.12213}{{\ttfamily
  arXiv:2006.12213 [hep-th]}}.

\bibitem{Bose:2020cod}
A.~Bose, A.~Sinha, and S.~S. Tiwari, ``{Selection rules for the S-Matrix
  bootstrap},'' \href{http://arxiv.org/abs/2011.07944}{{\ttfamily
  arXiv:2011.07944 [hep-th]}}.

\bibitem{Guerrieri:2021ivu}
A.~Guerrieri, J.~Penedones, and P.~Vieira, ``{Where is String Theory?},''
  \href{http://arxiv.org/abs/2102.02847}{{\ttfamily arXiv:2102.02847
  [hep-th]}}.

\bibitem{Cordova:2019lot}
L.~C\'ordova, Y.~He, M.~Kruczenski, and P.~Vieira, ``{The O(N) S-matrix
  Monolith},'' \href{http://dx.doi.org/10.1007/JHEP04(2020)142}{{\em JHEP}
  {\bfseries 04} (2020) 142}, \href{http://arxiv.org/abs/1909.06495}{{\ttfamily
  arXiv:1909.06495 [hep-th]}}.

\bibitem{Guerrieri:2020kcs}
A.~L. Guerrieri, A.~Homrich, and P.~Vieira, ``{Dual S-matrix bootstrap. Part I.
  2D theory},'' \href{http://dx.doi.org/10.1007/JHEP11(2020)084}{{\em JHEP}
  {\bfseries 11} (2020) 084}, \href{http://arxiv.org/abs/2008.02770}{{\ttfamily
  arXiv:2008.02770 [hep-th]}}.

\bibitem{He:2021eqn}
Y.~He and M.~Kruczenski, ``{S-matrix bootstrap in 3+1 dimensions:
  regularization and dual convex problem},''
  \href{http://arxiv.org/abs/2103.11484}{{\ttfamily arXiv:2103.11484
  [hep-th]}}.

\bibitem{GS2021}
A.~Guerrieri and A.~Sever, ``{In preparation},''
  \href{http://arxiv.org/abs/21XX.XXXXX}{{\ttfamily arXiv:21XX.XXXXX
  [hep-th]}}.

\bibitem{He:2018uxa}
Y.~He, A.~Irrgang, and M.~Kruczenski, ``{A note on the S-matrix bootstrap for
  the 2d O(N) bosonic model},''
  \href{http://dx.doi.org/10.1007/JHEP11(2018)093}{{\em JHEP} {\bfseries 11}
  (2018) 093}, \href{http://arxiv.org/abs/1805.02812}{{\ttfamily
  arXiv:1805.02812 [hep-th]}}.

\bibitem{Cordova:2018uop}
L.~C\'ordova and P.~Vieira, ``{Adding flavour to the S-matrix bootstrap},''
  \href{http://dx.doi.org/10.1007/JHEP12(2018)063}{{\em JHEP} {\bfseries 12}
  (2018) 063}, \href{http://arxiv.org/abs/1805.11143}{{\ttfamily
  arXiv:1805.11143 [hep-th]}}.

\bibitem{Paulos:2018fym}
M.~F. Paulos and Z.~Zheng, ``{Bounding scattering of charged particles in $1+1$
  dimensions},'' \href{http://dx.doi.org/10.1007/JHEP05(2020)145}{{\em JHEP}
  {\bfseries 05} (2020) 145}, \href{http://arxiv.org/abs/1805.11429}{{\ttfamily
  arXiv:1805.11429 [hep-th]}}.

\bibitem{Kruczenski:2020ujw}
M.~Kruczenski and H.~Murali, ``{The R-matrix bootstrap for the 2d O(N) bosonic
  model with a boundary},''
  \href{http://dx.doi.org/10.1007/JHEP04(2021)097}{{\em JHEP} {\bfseries 04}
  (2021) 097}, \href{http://arxiv.org/abs/2012.15576}{{\ttfamily
  arXiv:2012.15576 [hep-th]}}.

\bibitem{Rattazzi:2008pe}
R.~Rattazzi, V.~S. Rychkov, E.~Tonni, and A.~Vichi, ``{Bounding scalar operator
  dimensions in 4D CFT},''
  \href{http://dx.doi.org/10.1088/1126-6708/2008/12/031}{{\em JHEP} {\bfseries
  12} (2008) 031}, \href{http://arxiv.org/abs/0807.0004}{{\ttfamily
  arXiv:0807.0004 [hep-th]}}.

\bibitem{ElShowk:2012ht}
S.~El-Showk, M.~F. Paulos, D.~Poland, S.~Rychkov, D.~Simmons-Duffin, and
  A.~Vichi, ``{Solving the 3D Ising Model with the Conformal Bootstrap},''
  \href{http://dx.doi.org/10.1103/PhysRevD.86.025022}{{\em Phys. Rev. D}
  {\bfseries 86} (2012) 025022},
  \href{http://arxiv.org/abs/1203.6064}{{\ttfamily arXiv:1203.6064 [hep-th]}}.

\bibitem{El-Showk:2014dwa}
S.~El-Showk, M.~F. Paulos, D.~Poland, S.~Rychkov, D.~Simmons-Duffin, and
  A.~Vichi, ``{Solving the 3d Ising Model with the Conformal Bootstrap II.
  c-Minimization and Precise Critical Exponents},''
  \href{http://dx.doi.org/10.1007/s10955-014-1042-7}{{\em J. Stat. Phys.}
  {\bfseries 157} (2014) 869}, \href{http://arxiv.org/abs/1403.4545}{{\ttfamily
  arXiv:1403.4545 [hep-th]}}.

\bibitem{Kos:2016ysd}
F.~Kos, D.~Poland, D.~Simmons-Duffin, and A.~Vichi, ``{Precision Islands in the
  Ising and $O(N)$ Models},''
  \href{http://dx.doi.org/10.1007/JHEP08(2016)036}{{\em JHEP} {\bfseries 08}
  (2016) 036}, \href{http://arxiv.org/abs/1603.04436}{{\ttfamily
  arXiv:1603.04436 [hep-th]}}.

\bibitem{Luscher:1980ac}
M.~Luscher, ``{Symmetry Breaking Aspects of the Roughening Transition in Gauge
  Theories},'' \href{http://dx.doi.org/10.1016/0550-3213(81)90423-5}{{\em Nucl.
  Phys. B} {\bfseries 180} (1981) 317--329}.

\bibitem{Luscher:1980fr}
M.~Luscher, K.~Symanzik, and P.~Weisz, ``{Anomalies of the Free Loop Wave
  Equation in the WKB Approximation},''
  \href{http://dx.doi.org/10.1016/0550-3213(80)90009-7}{{\em Nucl. Phys. B}
  {\bfseries 173} (1980) 365}.

\bibitem{Dubovsky:2012sh}
S.~Dubovsky, R.~Flauger, and V.~Gorbenko, ``{Effective String Theory
  Revisited},'' \href{http://dx.doi.org/10.1007/JHEP09(2012)044}{{\em JHEP}
  {\bfseries 09} (2012) 044}, \href{http://arxiv.org/abs/1203.1054}{{\ttfamily
  arXiv:1203.1054 [hep-th]}}.

\bibitem{Aharony:2013ipa}
O.~Aharony and Z.~Komargodski, ``{The Effective Theory of Long Strings},''
  \href{http://dx.doi.org/10.1007/JHEP05(2013)118}{{\em JHEP} {\bfseries 05}
  (2013) 118}, \href{http://arxiv.org/abs/1302.6257}{{\ttfamily arXiv:1302.6257
  [hep-th]}}.

\bibitem{Caselle:2021eir}
M.~Caselle, ``{Effective string description of the confining flux tube at
  finite temperature},'' \href{http://arxiv.org/abs/2104.10486}{{\ttfamily
  arXiv:2104.10486 [hep-lat]}}.

\bibitem{Luscher:2004ib}
M.~Luscher and P.~Weisz, ``{String excitation energies in SU(N) gauge theories
  beyond the free-string approximation},''
  \href{http://dx.doi.org/10.1088/1126-6708/2004/07/014}{{\em JHEP} {\bfseries
  07} (2004) 014}, \href{http://arxiv.org/abs/hep-th/0406205}{{\ttfamily
  arXiv:hep-th/0406205}}.

\bibitem{Athenodorou:2010cs}
A.~Athenodorou, B.~Bringoltz, and M.~Teper, ``{Closed flux tubes and their
  string description in D=3+1 SU(N) gauge theories},''
  \href{http://dx.doi.org/10.1007/JHEP02(2011)030}{{\em JHEP} {\bfseries 02}
  (2011) 030}, \href{http://arxiv.org/abs/1007.4720}{{\ttfamily arXiv:1007.4720
  [hep-lat]}}.

\bibitem{Dubovsky:2013gi}
S.~Dubovsky, R.~Flauger, and V.~Gorbenko, ``{Evidence from Lattice Data for a
  New Particle on the Worldsheet of the QCD Flux Tube},''
  \href{http://dx.doi.org/10.1103/PhysRevLett.111.062006}{{\em Phys. Rev.
  Lett.} {\bfseries 111} no.~6, (2013) 062006},
  \href{http://arxiv.org/abs/1301.2325}{{\ttfamily arXiv:1301.2325 [hep-th]}}.

\bibitem{Aharony:2009gg}
O.~Aharony and E.~Karzbrun, ``{On the effective action of confining strings},''
  \href{http://dx.doi.org/10.1088/1126-6708/2009/06/012}{{\em JHEP} {\bfseries
  06} (2009) 012}, \href{http://arxiv.org/abs/0903.1927}{{\ttfamily
  arXiv:0903.1927 [hep-th]}}.

\bibitem{Aharony:2010cx}
O.~Aharony and M.~Field, ``{On the effective theory of long open strings},''
  \href{http://dx.doi.org/10.1007/JHEP01(2011)065}{{\em JHEP} {\bfseries 01}
  (2011) 065}, \href{http://arxiv.org/abs/1008.2636}{{\ttfamily arXiv:1008.2636
  [hep-th]}}.

\bibitem{Aharony:2011gb}
O.~Aharony and M.~Dodelson, ``{Effective String Theory and Nonlinear Lorentz
  Invariance},'' \href{http://dx.doi.org/10.1007/JHEP02(2012)008}{{\em JHEP}
  {\bfseries 02} (2012) 008}, \href{http://arxiv.org/abs/1111.5758}{{\ttfamily
  arXiv:1111.5758 [hep-th]}}.

\bibitem{Conkey:2016qju}
P.~Conkey and S.~Dubovsky, ``{Four Loop Scattering in the Nambu-Goto Theory},''
  \href{http://dx.doi.org/10.1007/JHEP05(2016)071}{{\em JHEP} {\bfseries 05}
  (2016) 071}, \href{http://arxiv.org/abs/1603.00719}{{\ttfamily
  arXiv:1603.00719 [hep-th]}}.

\bibitem{Polchinski:1991ax}
J.~Polchinski and A.~Strominger, ``{Effective string theory},''
  \href{http://dx.doi.org/10.1103/PhysRevLett.67.1681}{{\em Phys. Rev. Lett.}
  {\bfseries 67} (1991) 1681--1684}.

\bibitem{Dubovsky:2014fma}
S.~Dubovsky, R.~Flauger, and V.~Gorbenko, ``{Flux Tube Spectra from Approximate
  Integrability at Low Energies},''
  \href{http://dx.doi.org/10.1134/S1063776115030188}{{\em J. Exp. Theor. Phys.}
  {\bfseries 120} (2015) 399--422},
  \href{http://arxiv.org/abs/1404.0037}{{\ttfamily arXiv:1404.0037 [hep-th]}}.

\bibitem{Teper:2009uf}
M.~Teper, ``{Large N and confining flux tubes as strings - a view from the
  lattice},'' {\em Acta Phys. Polon. B} {\bfseries 40} (2009) 3249--3320,
  \href{http://arxiv.org/abs/0912.3339}{{\ttfamily arXiv:0912.3339 [hep-lat]}}.

\bibitem{Zamolodchikov:1991vx}
A.~Zamolodchikov, ``{From tricritical Ising to critical Ising by thermodynamic
  Bethe ansatz},'' \href{http://dx.doi.org/10.1016/0550-3213(91)90423-U}{{\em
  Nucl. Phys. B} {\bfseries 358} (1991) 524--546}.

\bibitem{Chen:2018keo}
C.~Chen, P.~Conkey, S.~Dubovsky, and G.~Hern\'andez-Chifflet, ``{Undressing
  Confining Flux Tubes with $T\bar T$},''
  \href{http://dx.doi.org/10.1103/PhysRevD.98.114024}{{\em Phys. Rev. D}
  {\bfseries 98} no.~11, (2018) 114024},
  \href{http://arxiv.org/abs/1808.01339}{{\ttfamily arXiv:1808.01339
  [hep-th]}}.

\bibitem{Athenodorou:2016kpd}
A.~Athenodorou and M.~Teper, ``{Closed flux tubes in D = 2 + 1 SU(N ) gauge
  theories: dynamics and effective string description},''
  \href{http://dx.doi.org/10.1007/JHEP10(2016)093}{{\em JHEP} {\bfseries 10}
  (2016) 093}, \href{http://arxiv.org/abs/1602.07634}{{\ttfamily
  arXiv:1602.07634 [hep-lat]}}.

\bibitem{Haldar:2021rri}
P.~Haldar, A.~Sinha, and A.~Zahed, ``{Quantum field theory and the Bieberbach
  conjecture},'' \href{http://arxiv.org/abs/2103.12108}{{\ttfamily
  arXiv:2103.12108 [hep-th]}}.

\bibitem{cvx}
S.~Boyd and L.~Vandenberghe {\em {Convex Optimization, Cambridge Univ. Press
  (2004) }} .

\bibitem{Dubovsky:2015zey}
S.~Dubovsky and V.~Gorbenko, ``{Towards a Theory of the QCD String},''
  \href{http://dx.doi.org/10.1007/JHEP02(2016)022}{{\em JHEP} {\bfseries 02}
  (2016) 022}, \href{http://arxiv.org/abs/1511.01908}{{\ttfamily
  arXiv:1511.01908 [hep-th]}}.

\bibitem{Athenodorou:2017cmw}
A.~Athenodorou and M.~Teper, ``{On the mass of the world-sheet 'axion' in
  $SU(N)$ gauge theories in 3$+$1 dimensions},''
  \href{http://dx.doi.org/10.1016/j.physletb.2017.05.082}{{\em Phys. Lett. B}
  {\bfseries 771} (2017) 408--414},
  \href{http://arxiv.org/abs/1702.03717}{{\ttfamily arXiv:1702.03717
  [hep-lat]}}.

\bibitem{Dubovsky:2008bd}
S.~Dubovsky and S.~Sibiryakov, ``{Superluminal Travel Made Possible (in two
  dimensions)},'' \href{http://dx.doi.org/10.1088/1126-6708/2008/12/092}{{\em
  JHEP} {\bfseries 12} (2008) 092},
  \href{http://arxiv.org/abs/0806.1534}{{\ttfamily arXiv:0806.1534 [hep-th]}}.

\bibitem{Cooper:2013ffa}
P.~Cooper, S.~Dubovsky, and A.~Mohsen, ``{Ultraviolet complete
  Lorentz-invariant theory with superluminal signal propagation},''
  \href{http://dx.doi.org/10.1103/PhysRevD.89.084044}{{\em Phys. Rev. D}
  {\bfseries 89} no.~8, (2014) 084044},
  \href{http://arxiv.org/abs/1312.2021}{{\ttfamily arXiv:1312.2021 [hep-th]}}.

\bibitem{Doroud:2018szp}
N.~Doroud and J.~Elias~Mir\'o, ``{S-matrix bootstrap for resonances},''
  \href{http://dx.doi.org/10.1007/JHEP09(2018)052}{{\em JHEP} {\bfseries 09}
  (2018) 052}, \href{http://arxiv.org/abs/1804.04376}{{\ttfamily
  arXiv:1804.04376 [hep-th]}}.

\bibitem{Conkey:2019blu}
P.~Conkey, S.~Dubovsky, and M.~Teper, ``{Glueball spins in $D = 3$
  Yang-Mills},'' \href{http://dx.doi.org/10.1007/JHEP10(2019)175}{{\em JHEP}
  {\bfseries 10} (2019) 175}, \href{http://arxiv.org/abs/1909.07430}{{\ttfamily
  arXiv:1909.07430 [hep-lat]}}.

\bibitem{Cooper:2014noa}
P.~Cooper, S.~Dubovsky, V.~Gorbenko, A.~Mohsen, and S.~Storace, ``{Looking for
  Integrability on the Worldsheet of Confining Strings},''
  \href{http://dx.doi.org/10.1007/JHEP04(2015)127}{{\em JHEP} {\bfseries 04}
  (2015) 127}, \href{http://arxiv.org/abs/1411.0703}{{\ttfamily arXiv:1411.0703
  [hep-th]}}.

\bibitem{Bercini:2019vme}
C.~Bercini, M.~Fabri, A.~Homrich, and P.~Vieira, ``{S-matrix bootstrap:
  Supersymmetry, $Z_2$, and $Z_4$ symmetry},''
  \href{http://dx.doi.org/10.1103/PhysRevD.101.045022}{{\em Phys. Rev. D}
  {\bfseries 101} no.~4, (2020) 045022},
  \href{http://arxiv.org/abs/1909.06453}{{\ttfamily arXiv:1909.06453
  [hep-th]}}.

\end{thebibliography}\endgroup
\bibliographystyle{utphys}

\end{document}